\documentclass[pra,amsmath,amssymb,twocolumn,superscriptaddress,footinbib]{revtex4-2}
\usepackage{amsmath}
\usepackage{amssymb}
\usepackage{amstext}
\usepackage{amsfonts}
\usepackage{amsxtra}
\usepackage{bm}
\usepackage{soul}
\usepackage{epstopdf}
\usepackage{xcolor}
\usepackage{graphicx}
\usepackage{marvosym}
\usepackage{wasysym}
\usepackage{dsfont}
\usepackage{upgreek}

\usepackage{graphicx}
\usepackage{array}
\usepackage{lipsum}  

\graphicspath{./Figures}
 \usepackage{booktabs}

\renewcommand{\vec}[1]{\mathbf{#1}}

\definecolor{applegreen}{rgb}{0.55, 0.71, 0.0}
\definecolor{byzantine}{rgb}{0.74, 0.2, 0.64}
\definecolor{dark green}{RGB}{0 150 0}

\makeatletter
\let\cat@comma@active\@empty
\makeatother

\DeclareSymbolFont{usualmathcal}{OMS}{cmsy}{m}{n}
\DeclareSymbolFontAlphabet{\mathcal}{usualmathcal}

\begin{document}

\title{Morphological false-vacuum decay in dipolar supersolids}
	
	\author{Wyatt Kirkby}
	\affiliation{Kirchhoff-Institut f\"ur Physik,
		Universit\"at Heidelberg,
		Im~Neuenheimer~Feld~227,
		69120~Heidelberg, Germany}
	\affiliation{Physikalisches Institut,
		Universit\"at Heidelberg,
		Im~Neuenheimer~Feld~226,
		69120~Heidelberg, Germany}

    \author{Lauriane Chomaz}
	\affiliation{Physikalisches Institut,
		Universit\"at Heidelberg,
		Im~Neuenheimer~Feld~226,
		69120~Heidelberg, Germany}

    \author{Thomas Gasenzer}
	\affiliation{Kirchhoff-Institut f\"ur Physik,
		Universit\"at Heidelberg,
		Im~Neuenheimer~Feld~227,
		69120~Heidelberg, Germany}

	%==============================================================================
	%==============================================================================
	\begin{abstract}
    False-vacuum decay between two morphologically distinct supersolid phases via bubble nucleation is studied in a uniform dipolar gas confined to the plane. Starting from a metastable honeycomb state, the formation of stripe phase domains is simulated numerically by means of a stochastic projected extended Gross-Pitaevskii equation. The speed of bubble growth is analyzed in relation to the multiple speeds of sound of the supersolid, and is found to be set by the slowest of these sounds. The vacuum decay rate is numerically extracted and compared against a minimal effective model for the Coleman bounce solution connecting the two supersolid orders. Our results establish dipolar supersolids as a novel and versatile platform for studying false-vacuum decay. This setting offers a rich structure of metastable states and collective excitations that come into play in the decay. Furthermore, here, in contrast to previous studies, bubble formation occurs directly in the real-space density and can be probed with \textit{in situ} imaging. 
    \end{abstract}
	\date{\today}
	
	\maketitle

	%==============================================================================
	%==============================================================================
    \section{Introduction}
    In the scenario proposed in 1977, Coleman describes the theory of false-vacuum decay (FVD) of a relativistic scalar field \cite{Coleman1977FateI,Callan1977FateII,Altland2010Book}, which tunnels from a metastable state through an energy barrier to a lower-energy configuration. Decay of the false vacuum to the absolute minimum energy configuration is seeded locally through fluctuations, typically thermal or quantum in nature, nucleating `bubbles' of the true vacuum that, once a critical size is achieved, expand at a speed limited by causality. The study of FVD and related mechanisms is important in a cosmological context, particularly in the context of the early universe, its ultimate fate, and hypothetical mechanisms for universe creation \cite{Farhi1990Is}. There has been renewed interest in the context of many-body physics and quantum simulation, where the FVD rate, i.e., the exponential suppression of the survival probability of a metastable state in time, can be directly probed. Such discontinuous phase transitions provide a natural arena for studying FVD, due to the typical multi-minimum structure of the resulting free-energy landscape. 

    %{\color{red}The FVD rate per unit volume can be estimated at leading order by calculating the so-called \textit{bounce} trajectory in imaginary time, corresponding to a classical trajectory in an inverted energy landscape, starting at and returning to false vacuum.}

    The versatility and accessibility of ultracold quantum gases has recently made them a tool of choice as an analog simulator of a wide range of many-body phenomena, including false-vacuum decay \cite{Fialko2015Fate,Fialko2017Universe}. Extensive theoretical work on possible scenarios has been explored, including examination of bubble-formation scenarios for ultracold gases \cite{Opanchuk2013Quantum,Braden2018Towards,Braden2019Nonlinear,Braden2019New,Braden2022Erratum,Billam2019SimulatingSeeded,Billam2020Simulating,Ng2021Fate,Billam2021SimulatingII,Billam2022Falsevacuum,Billam2023Bubble,Braden2023Mass,Jenkins2024Analog,Jenkins2024Generalized,Jenkins2025Bubbles,Brown2025Mitigating,Sivasankar2025Temperature}, as well as in quantum spin chains \cite{Lagnese2021False,Rutkevich1999Decay,Sinha2021Nonadiabatic,Pomponio2022Bloch,Milsted2022Collisions,Lencses2022Variations,Maki2023Monte,Darbha2024False,Lagnese2024Detecting,Pomponio2025Confinement,Yin2025Theory}. Tunneling probabilities, times, and bubble formation have been simulated in the Ising model using a quantum annealer \cite{Abel2021QuantumFieldTheoretic,Vodeb2025Stirring}, and FVD has also been observed in systems of neutral atoms in optical lattices \cite{Zhu2024Probing}, as well as with trapped ions \cite{Luo2025Quantum}. In a cold-gas context, FVD via bubble nucleation has been recently observed for the first time in a ferromagnetic superfluid \cite{Zenesini2024False}, where a finite-temperature approach to the instanton (bounce) trajectory in imaginary time was found to capture the dependence of decay rates with respect to external parameters.

    The supersolid phase, a state which supports superfluidity along with regular crystal order in the density (i.e. simultaneous off- and on-diagonal long-range order in the density matrix) \cite{Penrose1956B,Gross1957Unified,Gross1958Classical,Andreev1969Quantum,Leggett1970Can,Boninsegni2012Colloquium}, has recently been realized in experiments with atomic quantum gases \cite{Leonard2017Supersolid,li2017stripe,Bersano2019Experimental}, and in particular strongly magnetic dipolar atoms \cite{Tanzi2019Observation,Boettcher2019Transient,Chomaz2019LongLived,Chomaz2022Dipolar}. In the latter, the transition between ordinary superfluid and supersolid is determined by the competition between short- and long-range interactions, with the ratio typically being tuned by using atomic Feshbach resonances. To date, supersolid droplet states have been realized in both one- and two-dimensional arrangements \cite{Norcia2021Twodimensional,Bland2022TwoDimensional}. Furthermore, recent progress with polar molecules has resulted in similar droplet states \cite{Zhang2025Observation} and also opened the door to the possible creation of similar states using molecules with extremely strong magnetic moments. Dipolar supersolids in two dimensions offer a rich platform to study vacuum decay since they possess a rich phase diagram with multiple first-order phase transitions, across which large regions of bistability between supersolid orders can exist. 

    Up to now, most studies of FVD in ultracold gases focus on the transition between internal degrees of freedom (e.g. spin states, phase angles). In this paper, we consider instead a single-component system undergoing a morphological change between different two-dimensional supersolid density structures. The supersolid platform offers the qualitatively new feature of bubble nucleation directly in the real-space density of an atomic gas, which is experimentally observable using \textit{in situ} imaging techniques, see e.g.\,\cite{Boettcher2019Transient,Sohmen2021Birth,Norcia2021Twodimensional,Bland2022TwoDimensional}. The change in real-space structure between two crystalline orders is therefore reminiscent of transitions between competing metastable structures seen in other condensed-matter contexts, such as polar-domain patterns in ferroic materials \cite{Nahas2020Topology}, phases of paramagnetic colloidal crystals \cite{Alert2016Formation}, and skyrmion lattice configurations \cite{Oike2015Interplay,Karube2016Robust}. Conceptually similar changes can also be observed in charge-density-wave configurations of layered compounds \cite{Stojchevska2014Ultrafast,Vaskivskyi2015Controlling,Ma2016Metallic,Gerasimenko2019Intertwined}. Furthermore, the complex landscape of supersolid orders provides a highly tunable arena for different (meta)stable configurations and collective excitation modes, which both play a crucial role in FVD physics. In particular, supersolid states display multiple sound modes and the question arises of which sound velocity sets the speed at which the surface of the true vacuum bubble expands. 
    
    In Section~\ref{sec:system}, we present our model for a dipolar gas confined to a two-dimensional plane, as well as the ground-state phase diagram. We point out where one can identify true and false vacua in the supersolid system, and give the equation of motion for our dynamical simulations. Sec.~\ref{sec:BubbleFormation} shows our numerical simulations demonstrating the dynamical formation of true vacuum bubbles, and analyze the speed of bubble growth. In Sec.~\ref{sec:Rates}, we demonstrate how one can derive a minimal analytical model of supersolid false-vacuum decay, and compare the resulting instanton calculations to the numerically extracted decay rates.
    
	%==============================================================================
	%==============================================================================
    \section{System}
    \label{sec:system}
    
	%==============================================================================
    \subsection{Ultracold dipolar gas: Model}
    \label{subsec:model}

    In this paper, we focus on the two-dimensional supersolid phases of a polarized strongly dipolar gas of ${}^{164}\mathrm{Dy}$. We take the system to be harmonically trapped along the axial polarization direction (hereby chosen to be the $z$-direction), and untrapped in the $x-y$ plane. We consider a functional-integral approach, in which we assume that the field describing the bosonic gas is separable along the trapped direction \cite{Zhang2019Supersolidity,Schmidt2022Selfbound,Ripley2023Twodimensional,Lima2025Supersolid}, 
    \begin{equation}
        \Psi(\mathbf{r})=\psi(\mathbf{x},t)\varphi(z)\;,
    \end{equation}
    where $\mathbf{x}=(x,y)$. The axial field is described by a Gaussian profile,
    \begin{equation}
        \varphi(z)=\frac{1}{\sqrt{\ell}\pi^{1/4}}\mathrm{e}^{-z^2/(2\ell^2)}\;,
    \end{equation}
    where $\ell$ is a variational width determined by a minimization of the energy, subject to $\int\mathrm{d}z|\varphi(z)|^2=1$. The full 3D equations of motion can therefore be dimensionally reduced. This dimensional reduction has been shown to have generally good agreement with the full 3D system \cite{Lee2022Stability,Ripley2023Twodimensional}, outperforming typical quasi-2D theory where the width of the condensate $\ell$ would instead match the harmonic trap.

    The energy of the system is given by the effective classical Hamiltonian functional.
    Within a mean-field approximation, extended perturbatively by a quantum-fluctuation correction, this functional takes the form of the Hamiltonian of the extended Gross-Pitaevskii equation (eGPE) \cite{Pedri2005TwoDimensional,Baillie2015General,Waechtler2016Groundstate,Ripley2023Twodimensional,Zhang2024Metastable},
    \begin{align}
        &E_{\mathrm{eGP}}\;=\int_{\mathrm{uc}} \mathrm{d}\mathbf{x}\;\Biggl[\frac{\hbar^2}{2m}|\nabla\psi|^2+\mathcal{E}_z|\psi(\mathbf{x})|^2+\frac{\tilde{g}}{2}|\psi(\mathbf{x})|^4
        \nonumber\\
        &\;+\frac{1}{2}\int\mathrm{d}\mathbf{x}'|\psi(\mathbf{x})|^2 U^{2\mathrm{D}}(\mathbf{x}-\mathbf{x}')|\psi(\mathbf{x}')|^2
        +\frac{2\tilde{\gamma}_Q}{5}|\psi(\mathbf{x})|^5\Biggr]\,,
        \label{eq:2DEnergy}
    \end{align}
    where `uc' denotes the volume of a unit cell in a spatially periodic system. 
    $\mathcal{E}_z=m\omega_z^2\ell^2/4+{\hbar^2}/({4m\ell^2})$ describes the transverse trapping, where $\omega_z$ is the harmonic trapping frequency in the $z$-direction.  
    The reduced contact interaction strength is $\tilde{g}=g/(2\sqrt{2\pi}\ell)$, where the 3D strength is $g=4\pi\hbar^2a_\mathrm{s}/m$ for an atomic $s$-wave scattering length $a_\mathrm{s}$. The bare two-dimensional dipole-dipole interaction (DDI) is known explicitly in Fourier space \cite{Fischer2006Stability,Ticknor2011Anisotropic,Baillie2015General},
    \begin{equation}
        \tilde{U}^{2\mathrm{D}}(\mathbf{k})=\frac{2\sqrt{2\pi}\mu_0\mu_m^2}{3\ell}\left[2-3\sqrt{\frac{\pi}{2}}k\ell\mathrm{e}^{\frac{k^2\ell^2}{2}}\mathrm{erfc}\left(\frac{k\ell}{\sqrt{2}}\right)\right]\,,
    \end{equation}
    where $\mu_m=9.93\,\mu_{\mathrm{B}}$ is the magnetic moment of dysprosium ($\mu_{\mathrm{B}}$ is the Bohr magneton), and $\mu_0$ is the vacuum permeability. Within a mean-field approximation, the attractive interactions cause the trapped cloud to undergo a collapse for $a_{\mathrm{dd}}>a_\mathrm{s}$ where the dipolar length is $a_{\mathrm{dd}}=\mu_0\mu_m^2/(12\pi\hbar^2)$. Yet, for sufficiently strong dipolar interactions, beyond-mean-field quantum fluctuations \cite{Schuetzhold2006Meanfield,Lima2011Quantum,Lima2012Beyond,Chomaz2026Quantum} may stabilize the system, then forming droplet or supersolid phases~\cite{Waechtler2016Quantum,Waechtler2016Groundstate,Bisset2016Groundstate}. This is included in the effective extended mean-field energy functional, Eq.~\eqref{eq:2DEnergy}, via $\tilde{\gamma}_Q=(2/5)^{3/2}\pi^{-3/4}\ell^{-3/2}\gamma_{Q}$, where  
    %$\frac{2\gamma_{Q}}{5\pi^{3/4}\ell^{3/2}}\sqrt{\frac{2}{5}}$, and,  
    %For dipolar length given by $a_{\mathrm{dd}}=\mu_0\mu_m^2/(12\pi\hbar^2)$, the mean-field theory is susceptible to collapse when $a_{\mathrm{dd}}>a_\mathrm{s}$. For sufficiently strong dipolar interactions, the supersolid phase remains stable due to the presence of beyond-mean-field quantum fluctuations \cite{Schuetzhold2006Meanfield,Lima2011Quantum}, with $\tilde{\gamma}_Q=\frac{2\gamma_{Q}}{5\pi^{3/4}\ell^{3/2}}\sqrt{\frac{2}{5}}$, and,
    \begin{equation}
        \gamma_Q=\frac{128\sqrt{\pi}\hbar^2a_\mathrm{s}^{5/2}}{3m}\left[1+\frac{3}{2}\left(\frac{a_{\mathrm{dd}}}{a_\mathrm{s}}\right)^2\right]\,.
    \end{equation}
    
	%==============================================================================
    \subsection{Ground state properties}
    \label{subsec:groundstate}

    Within the extended mean-field approximation chosen above, the ground-state properties of the system are determined by the eGPE,
    \begin{equation}
        \mathrm{i}\hbar\frac{\partial \psi}{\partial t}=H_{\mathrm{eGP}}[\psi]\psi\,,\label{eq:2DGPE}
    \end{equation}
    where the dimensionally reduced non-linear Gross-Pitaevskii operator is,
\begin{align}
        H_{\mathrm{eGP}}=&\;-\frac{\hbar^2}{2m}\nabla^2+\mathcal{E}_z+\tilde{g}|\psi(\mathbf{x},t)|^2\nonumber+\tilde{\gamma}_{Q}|\psi(\mathbf{x},t)|^3\\&\;+\int\mathrm{d}\mathbf{x}'U^{2\mathrm{D}}(\mathbf{x}-\mathbf{x}')|\psi(\mathbf{x}',t)|^2-\mu\;,
    \end{align}
    with chemical potential $\mu$. At a given point in parameter space, the ground-state and metastable-state field configurations can be determined by solving Eq.~\eqref{eq:2DGPE} in imaginary time with initial biases corresponding to the desired geometry. Due to its nature as a convolution, the long-range dipolar interaction is efficiently evaluated in Fourier space. However, numerical evaluation of any long-range potential on a finite computational grid in momentum space introduces unphysical alias Fourier copies that self-interact, necessitating modified dipole-dipole interaction potentials with long-distance cutoffs (see e.g. \cite{Ronen2006Bogoliubov,Lu2010Spatial,Antoine2018Preconditioned,Lee2021Numerical,Smith2023Supersolidity}) for accurate results in finite (trapped) systems. In our untrapped system, however, we leverage the self-interaction introduced by the Fourier aliasing to reach the thermodynamic limit for ground-state calculations, and periodic boundary conditions for dynamic simulations, by using the bare dipole-dipole potential with no cutoff.
    
    In this paper we consider the mean atomic density of the 2D confined gas, conveniently defined as
    \begin{equation}
        \bar{n}=\frac{2}{\sqrt{3}L_{\mathrm{uc}}^2}
        \int_{\mathrm{uc}}\mathrm{d}\mathbf{x}\,|\psi(\mathbf{x})|^2\,,
        \label{eq:meandensity}
    \end{equation}
    where $L_{\mathrm{uc}}$ is the side length of a rhombic unit cell of the dipolar supersolid. In general, increasing the mean density results in a greater contribution to the interaction energy of the system albeit in a highly non-trivial way, due to the simultaneous competition of contact, dipolar, and LHY contributions.
    
    In order to calculate ground states, we fix $L_{\mathrm{uc}}$ and evaluate Eq.~\eqref{eq:2DGPE} in imaginary time. We then modify $\ell$ to minimize the energy per particle (Eq.\eqref{eq:2DEnergy} divided by $N$) and iteratively evaluate Eq.~\eqref{eq:2DGPE} until the combined solution of $\psi$ and $\ell$ minimizes the energy per particle within a tolerance of $\Delta \ell/\ell\leq 10^{-4}$ and with a normalized residual $r\equiv \int\mathrm{d}\mathbf{x}|H_{\mathrm{eGP}}\psi|/\int\mathrm{d}\mathbf{x} |\mu\psi|$ with a tolerance\footnote{Normally, this level of convergence may not be necessary, however a strict tolerance on convergence on $\psi$ is necessary for accurate low-$\mathbf{k}$ Bogoliubov-de Gennes modes.} of $r\leq 10^{-9}$ . The ground-state unit cell for the states considered in this paper can then be determined by repeating this procedure over by increasing or decreasing $L_{\mathrm{uc}}$ until Eq.~\eqref{eq:2DEnergy} is minimized, where again we choose a tolerance of $\Delta L_{\mathrm{uc}}/L_{\mathrm{uc}}\leq 10^{-4}$. In order to accelerate convergence, we select a rhombus unit cell with lattice vectors at an angle of $\pi/3$, since square grids prohibit single-site equilateral unit cells. Initial biases with the appropriate supersolid symmetry are introduced to the imaginary-time evolution to speed up convergence and to ensure metastable states can also be found. These biases correspond to cosine-modulated \textit{ans\"{a}tze} [given in Eqs.~\eqref{eq:ansatz} and \eqref{eq:app:ansatz}]. The (meta)stability of each state can further be confirmed by performing a Bogoliubov-de Gennes analysis of linear excitations on the resulting states and verifying that no instabilities are present, see Appendix~\ref{app:BogTheory}. 

	%==================================================
    \begin{figure}
        \centering
        \includegraphics[width=1.0\linewidth]{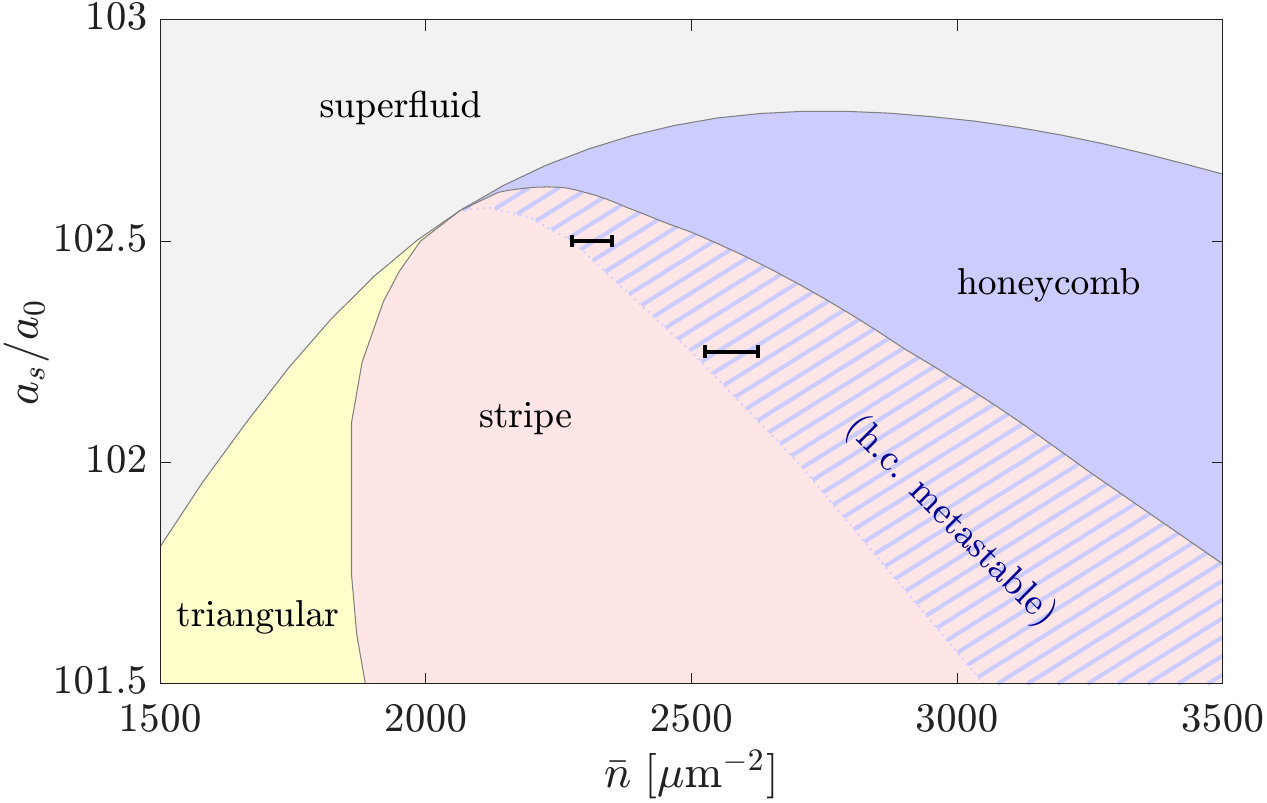}
        \caption{Phase diagram for dipolar supersolids in an extended 2D geometry, as a function of the $s$-wave scattering length $a_\mathrm{s}$ normalized by the Bohr radius $a_0$, and the mean atomic 2D planar density $\bar{n}$, defined as the total number of atoms per unit area via Eq.~\eqref{eq:meandensity}. The dipolar length of dysprosium gives $a_{\mathrm{dd}}\approx 131a_0$. As described in the main text, the phases are calculated by solving Eq.~\eqref{eq:2DGPE} in imaginary time, then minimizing the energy per particle via \eqref{eq:2DEnergy} as a function of unit cell size. The hatched region corresponds to a region of bistability, where the stripe phase is lower in energy, but the honeycomb (h.c.) is metastable. The density ranges at the two scattering lengths we examine in this paper, $a_\mathrm{s}=102.25\,a_0$ and $102.5\,a_0$, are highlighted.}
        \label{fig:PhaseDiagram}
    \end{figure}
	%==================================================

    The ground-state phase diagram for this system has been previously calculated \cite{Zhang2019Supersolidity,Ripley2023Twodimensional,Zhang2024Metastable,Lima2025Supersolid} and is shown in Fig.~\ref{fig:PhaseDiagram} for the relevant parameter ranges we consider here. We present the phase diagram as a function of the scattering length $a_\mathrm{s}$, which in experiment can be tuned by magnetic field strength through Feshbach resonances \cite{Chomaz2022Dipolar}, and mean 2D density, which can be tuned by varying the particle number. At large $a_\mathrm{s}$, the system is an unmodulated superfluid, which transitions to a modulated supersolid phase when decreasing $a_\mathrm{s}$. With increasing mean density, the density modulation changes from a triangular arrangement, through a striped one, to a honeycomb form. The triangular phase corresponds to high-density droplets in an equilateral arrangement with a weaker superfluid background, while the honeycomb is essentially the inverse: an array of `dimples' in a stronger superfluid background. When the magnetic moment is polarized along the direction of strong confinement (the case we consider here), transitions between different phases are all first-order except at a single critical point where the transitions meet, at which the transition is second-order. Including dipolar tilt breaks azimuthal symmetry by introducing a preferential direction, but also leads to greater complexity in the phase diagram, with a mixture of first- and second-order transitions between most states \cite{Lima2025Supersolid}. We will not be considering any dipolar tilts in the remainder of the manuscript.
    
    It has been demonstrated \cite{Blakie2025Dirac} that in the stripe phase, there is a region where the honeycomb state remains metastable and supports a nonzero transverse shear sound mode. Deeper in the stripe regime, either by lowering the scattering length or decreasing the density, eventually the transverse sound mode drops to zero, resulting in a shear instability of the honeycomb crystal and a loss of bistability. Practically, we calculate the metastable honeycomb states using the procedure described above, and for each state we numerically solve the Bogoliubov-de Gennes equations \eqref{eq:BdGg2}-\eqref{eq:BdGf2} to extract the respective modes. For the honeycomb states, we track the low-$\mathbf{k}$ modes and identify where the shear-mode branch softens and eventually develops imaginary values of $\omega$, indicating the shear-instability boundary. This boundary is then shown as a function of mean density $\bar{n}$ and scattering length $a_s$ in Fig.~\ref{fig:PhaseDiagram}.
    
    Other complex metastable phases have also been found, including hexagonal droplet arrays and ring-like patterns \cite{Zhang2024Metastable}. In Fig.~\ref{fig:PhaseDiagram}, we show the region where the stripe phase is the ground state, but the honeycomb state is metastable (no shear instability) as a hatched area. It is the metastability of the honeycomb phase that we exploit in our study, although in principle other regions of the phase diagram where metastable phases exist could be used to study FVD in a similar context. 

    The honeycomb phase maintains a strong superfluid fraction across the entire phase region, due to the fact that all unit cells are continuously connected with superfluid. This is in contrast with the triangular phase, which has a superfluid fraction that decays relatively rapidly as the $s$-wave scattering length is decreased and each individual droplet forming the sites of the triangular structure becomes more isolated. For this reason, we can consider the entire honeycomb phase as supersolid, independent of density and scattering length. The stripe phase is an intermediate case, where the superfluid fraction must be interpreted as a tensor \cite{Blakie2024Superfluid} since it is strongly directionally dependent. The stripes maintain a superfluid fraction of unity along their orientation due to the preserved translational symmetry, but has a transverse superfluid fraction that drops with decreasing scattering length. In the region of the phase diagram we consider here, the stripe phase maintains a strong superfluid fraction ($\gtrsim 0.7$) in the ground state even along the modulated direction \cite{Ripley2023Twodimensional}.

	%==============================================================================
    \subsection{Dynamics}
    \label{subsec:dynamics}
    In order to simulate false-vacuum decay via bubble nucleation, we perform finite-temperature simulations using the stochastic projected eGPE (SPeGPE) \cite{Rooney2012Stochastic,Rooney2014a,Bland2022TwoDimensional},
    \begin{equation}
        \mathrm{i}\hbar\frac{\partial \psi}{\partial t}=\hat{\mathcal{P}}\left\{(1-\mathrm{i}\gamma)H_{\mathrm{eGP}}[\psi]\psi+\eta\right\}\;,\label{eq:SPGPE}
    \end{equation}
    with stochastic noise subject to $\langle\eta^*(\mathbf{x},t)\eta(\mathbf{x}',t)\rangle=2\hbar\gamma k_\text{B}T\delta(\mathbf{x}-\mathbf{x}')\delta(t-t')$. For this study, we select a temperature of $T=5\,\mathrm{nK}$ (see below and Appendix~\ref{app:Temperature} for a more detailed discussion of the effects of temperature). The dissipation parameter $\gamma$ does not have a known fixed value that matches experimental observations directly for a planar dipolar gas \cite{Linscott2014Thermally}. Although the choice of $\gamma$ may affect the exact time at which bubble formation occurs, since we are comparing all results at a particular fixed value of $\gamma$, we expect the general scaling to hold regardless of which value we choose. We select $\gamma=1.5\times 10^{-2}$, approximately in line with the values used in Refs.~\cite{Bland2022TwoDimensional,Billam2023Bubble}. Within SPeGPE theory, the system modes are subdivided into regions where the excitations are coherent (well-occupied [$N_{\mathrm{ex}}\gg$], low-energy) and incoherent (weakly occupied [$N_{\mathrm{ex}}\lesssim1$], high-energy) modes~\cite{Gardiner2003Stochastic,Blakie2008Dynamics,Rooney2012Stochastic,Rooney2014a}. The projection operator $\hat{\mathcal{P}}$ projects into the coherent region and formally should be taken to be defined by the Bogoliubov spectrum of the modulated system. However, this would be very challenging dynamically, and so we approximate the implementation by removing free-particle $k$-modes with energy larger than $E_{\mathrm{cut}}=2\mu$. Similar studies \cite{Billam2022Falsevacuum} implement the projector using a grid-dependent cutoff, but for our relatively high-resolution simulations ($k_{\mathrm{cut}}\sim k_{\mathrm{max}}/2$) this choice sets the mode occupation at the cutoff to be far too low to properly describe an incoherent bath. We elect to maintain a consistent cutoff across all simulations, where the mode occupations at the cutoff are $\sim0.1-0.5$ particles, following similar works performed at experimentally-realizable temperatures \cite{Bradley2008Bose,Rooney2013Persistentcurrent,McDonald2020Dynamics,Bland2022TwoDimensional}.
    
    %The projection operator $\hat{\mathcal{P}}$ removes modes with $E>2\mu$ \cite{Rooney2013Persistentcurrent,McDonald2020Dynamics,Bland2022TwoDimensional}, such that the modes above this cutoff are outside the range of coherently populated low-energy modes \cite{Gardiner2003Stochastic,Blakie2008Dynamics,Rooney2012a.PhysRevA.86.053634,Rooney2014a}. 
 
    In each dynamical simulation, we prepare the system by selecting the initial state to be in the metastable honeycomb configuration, with a size of 32 by 32 unit cells. We note that the differences in the variational parameter $\ell$ and the chemical potential $\mu$ between the false and true vacua are less than 0.1\%. Furthermore, while the single unit-cell size $L_{\mathrm{uc}}$ between the two states can differ by approximately $5\%$, performing a dynamical simulation on system sizes of $(32)^2$ unit cells ensures the system is able to relax to a state with an $L_{\mathrm{uc}}$ within $0.2\%$ of its `true' size. Experimentally, this initial state could be prepared by adiabatically crossing the first-order phase transition from the honeycomb phase into the metastable region of the stripe phase, akin to supercooling.

    Before we present our results, the effects of temperature on FVD should be addressed. In general, nonzero temperature can have a crucial role in seeding bubble formation, either by enhancing the underlying quantum fluctuations, or by triggering decay directly. Depending on the strength of the effects of these temperature fluctuations, one can continuously move from a `quantum' regime at low (but still nonzero) temperature, to `thermally assisted' tunneling at intermediate temperatures, and `classical' thermally activated tunneling at high temperatures \cite{Liu2002Quantum}. In each case, there exists a slightly modified instanton picture \cite{Linde1983Decay}; we later assess which of these instanton approaches applies to our system via a minimal model. From the perspective of thermally-enhanced quantum decay, one can picture that thermal fluctuations `push' the system higher in the free-energy landscape where the potential barrier may be thinner and easier to tunnel through, i.e., they essentially lower the depth of the occupied minimum in the effective energy landscape of field configurations. Later, we shall see that the quantum instanton model is sufficient to qualitatively model the tunneling at the simulated temperature (5nK), and so we estimate that we are dominated by quantum tunneling effects with minimal thermal assistance. 

    The temperature for our SPeGPE simulations is chosen to remain as close as possible to the zero-temperature limit, while maintaining reasonable simulation times. If we instead use a zero-temperature eGPE with only quantum noise and no projection, the decay times (if decay is possible) become longer than reasonable simulations. Including only the projector, i.e. a dissipative mechanism, but no dynamical noise into the zero-temperature eGPE appears to be able to realize decay \cite{Billam2022Falsevacuum}, even at zero temperature. Using this approach, statistics would need to be collected instead via a truncated-Wigner framework, building in thermal and quantum noise into the initial condition according to the Wigner function. A comparison between these two independent methods of simulating FVD could prove to be interesting and useful, however it is beyond the scope of this paper. Given our approach, including the choice of $\gamma$ and $E_{\mathrm{cut}}$ and our variational-2D description, we highlight that our goal here is to present a new ultracold quantum gas platform for which morphological FVD is possible, and qualitatively describe trends as a function of standard control parameters of the dipolar supersolid system. We believe that a more precise calculation of decay rates, quench protocols, and benchmarking our SPeGPE parameters \cite{Bland2022TwoDimensional} in comparison with experimental setups will then provide a natural extension to this work.
    
    In our analysis of the FVD rate (see Sec.~\ref{sec:Rates}), we present the low-temperature limit described by the theory of quantum tunneling via an instanton bounce. We also present a higher-temperature case in App.~\ref{app:Temperature} in which we consider instead thermal activation of the tunneling, similar to other studies  \cite{Zenesini2024False,Sivasankar2025Temperature} that have included additional fitting parameters, such as the effective inverse temperature $\beta=1/k_\mathrm{B}T$ (not to be compared directly with the SPeGPE temperature). As we will see, the zero-temperature instanton theory with a single amplitude fit is already sufficient to reasonably well capture the qualitative trend for a range of densities considered in the main text.

     % We estimate that this is sufficiently cold for the instanton bounce calculation to at least qualitatively describe the decay, since the rate as a function of the density appears to be reasonably well captured without the need of additional inclusion of temperature directly into Eq.~\eqref{eq:GammaBounce}. 
    
    %With current experimental capabilities with highly magnetic atoms, it is likely that the temperature would be closer to $\sim 50\,\mathrm{nK}$, and the finite-temperature seeding of the decay may be needed in the theory. 

	%==================================================
    \begin{figure*}
        \centering
        \includegraphics[width=0.8\linewidth]{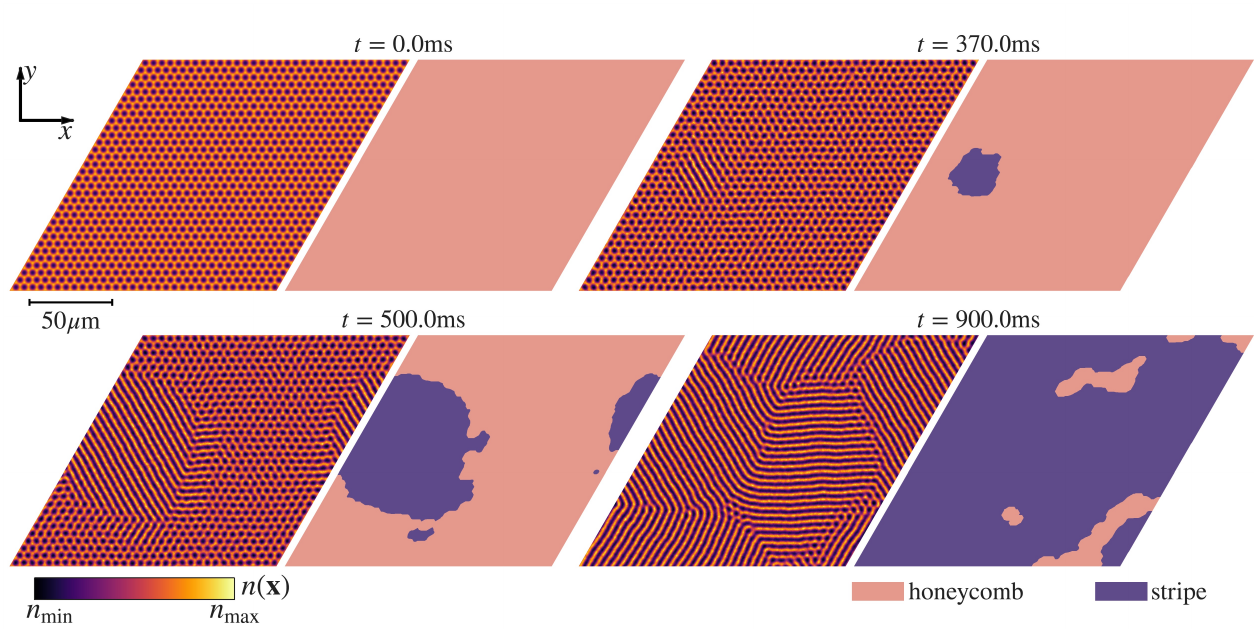}
        \caption{Bubble formation in dipolar supersolids. Starting from a metastable honeycomb state at $\bar{n}=2550\,\mu\mathrm{m}^{-2}$ and $a_\mathrm{s}=102.25\,a_0$, the decay to the stripe phase is locally triggered by some fluctuation. At each time step, the left panel shows the real-space density $n(\vec{x})=|\psi(\vec{x})|^2$, while the right panel shows the regions of honeycomb and stripe order in different color, as identified via the anisotropy parameter $\Xi$, cf. Eq.~\eqref{eq:Anisotropy} in Appendix \ref{app:RegionIdent}.}
        \label{fig:Bubbles}
    \end{figure*}
	%==================================================

	%==============================================================================
	%==============================================================================
    \section{Bubble formation}
    \label{sec:BubbleFormation}

	%==============================================================================
    \subsection{Bubble nucleation and growth}

    In Fig.~\ref{fig:Bubbles}, we show typical examples of morphological bubble formation we find in the two-dimensional supersolid system. Each snapshot in time shows the spatial density pattern $n(\vec{x})=|\psi(\vec{x})|^2$ in the respective left panel and the corresponding phase patterns in the right panel. The system remains in the initially developed honeycomb configuration until, at some point, fluctuations seed a perturbation large enough to induce a bubble with stripe arrangement, typically beginning with the merging of two or more honeycomb cells to form small stripes. The method used to automatically distinguish regions is described in App.~\ref{app:RegionIdent}.
    
    Although there is no externally imposed bias to the stripe direction due to the azimuthal symmetry of the DDI (transverse dipole orientation), they tend to align along one of the three symmetry axes of the initial honeycomb state. The initially small stripe bubble(s) then typically grow(s) in size and eventually merge together until covering nearly the full system. Since there is almost no energy cost of bending the stripes for our transverse dipole orientation, the long-time result of any given simulation can be therefore one of an essentially infinite amount of nearly-degenerate labyrinthine-like supersolid states \cite{Hertkorn2021Pattern}. Below the shear instability boundary, we observe no bubble formation and, instead, the entire crystal uniformly dissolves into the stripe phase (not shown).

	%==================================================
    \begin{figure}
        \centering
        \includegraphics[width=1.0\linewidth]{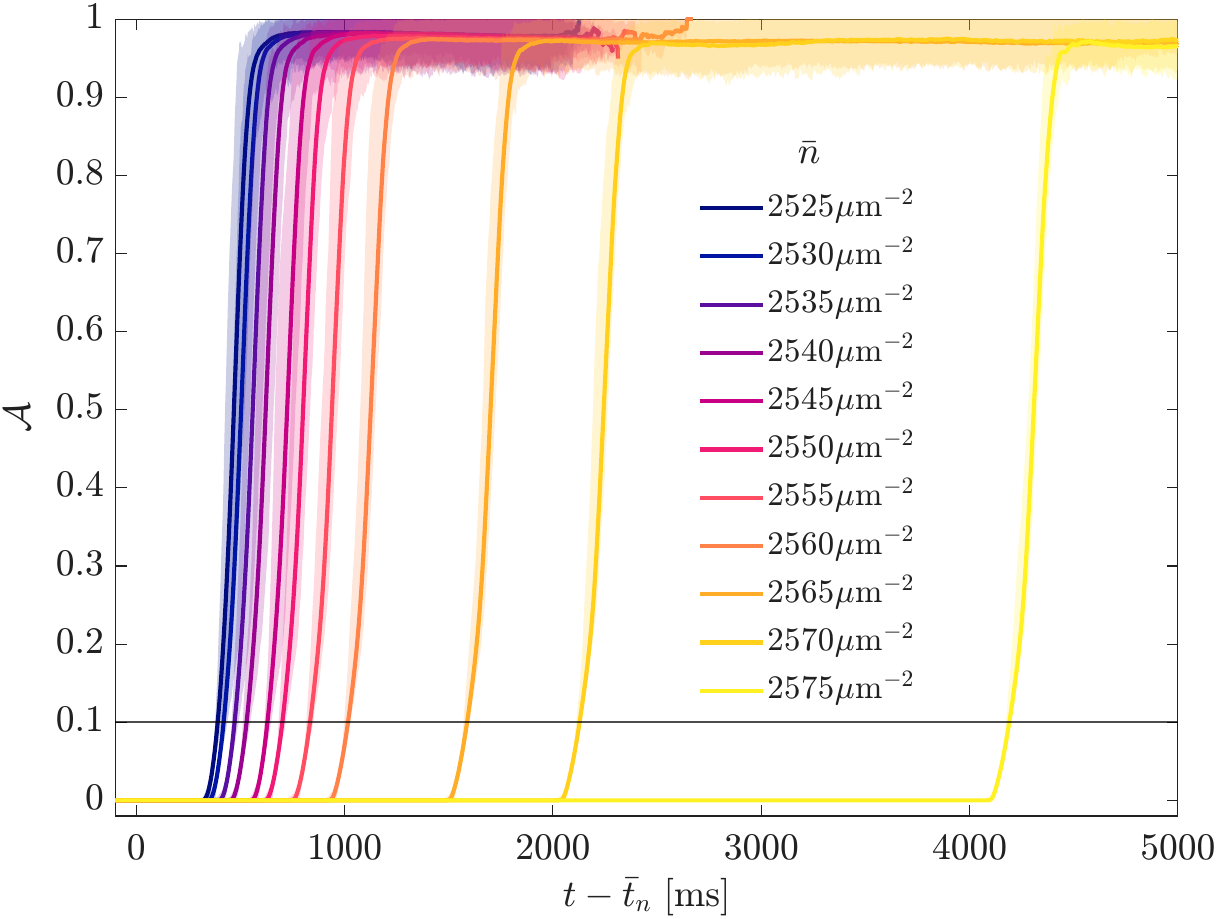}
        \caption{Total relative area of the true vacuum as a function of time, for scattering length $a_\mathrm{s}=102.25\,a_0$ and a range of average densities $\bar{n}$ defined in Eq.~\eqref{eq:meandensity}. Each curve is shifted to the average bubble nucleation time $\bar{t}_n$ at each respective density. The spread of all 250 trials (all shifted to the corresponding nucleation time) is shown as a shaded area with the mean shifted curve plotted as a solid line. The value $\mathcal{A}=0.1$ at which $\mathrm{v}$ is extracted is highlighted as a horizontal line.}
        \label{fig:BubbleAreas}
    \end{figure}
	%==================================================

    We show, in Fig.~\ref{fig:BubbleAreas}, the time dependence of the area of the striped region, relative to the total simulation area, averaged over 250 independent runs, for a set of 7 different 2D atomic mean densities $\bar{n}$, as defined by Eq.~\eqref{eq:meandensity}. Since bubble nucleation may occur randomly at any time, showing all 250 trajectories for each fixed density on the same graph makes the resulting plot inscrutable (see Fig.~\ref{fig:Areas_Unshifted} for a selection of raw trial data). In order to better portray the typical growth behaviour of any given trial on the same footing, we first shift all trajectories for a given density such that their individual nucleation times match the mean nucleation time $\bar{t}_n$, and then take the average to give the solid curves in Fig.~\ref{fig:BubbleAreas}. We observe that the mean nucleation time grows with density, while the growth rate after nucleation varies only weakly. We will analyze these behaviors further below. The shaded area is the spread of all 250 trials around the mean curve, showing that once nucleation begins, there is very little variation in growth rate at at any point until saturation. Small fluctuations in the growth rate likely occur due to multi-nucleation events. The spread of saturation values from $\mathcal{A}\approx 0.9$ to 1 occurs due to pockets of identified honeycomb configuration at the boundaries of regions of different stripe orientations. At these boundaries, the stripes can either smoothly bend or will maintain small superfluid rings that are identified as individual honeycomb cells (see Fig.~\ref{fig:Bubbles} at $t=900\,\mathrm{ms}$). The late-time tails of the averaged curves shown in Fig.~\ref{fig:BubbleAreas} are an artifact of shifting all of the individual trials, so that at late times each curve is averaged over progressively fewer trials. The focus of this paper will be on times well before saturation, and so these tails are not relevant to our results.

    To finish this section, we focus on bubble nucleation directly. In both FVD and when studying e.g. phase-ordering kinetics \cite{Bray1994Theory}, the interface energy also becomes a relevant quantity. The total interface energy depends on both the functional form in which the order parameter varies across the domain boundary and on the total surface area or length of the interface. In FVD, this energy cost is often balanced by the energy gain proportional to the bubble’s volume, which gives rise to a critical bubble size: small bubbles cost too much energy and therefore do not suffice to induce large-scale nucleation.

    \begin{figure*}
        \centering
        \includegraphics[width=0.75\linewidth]{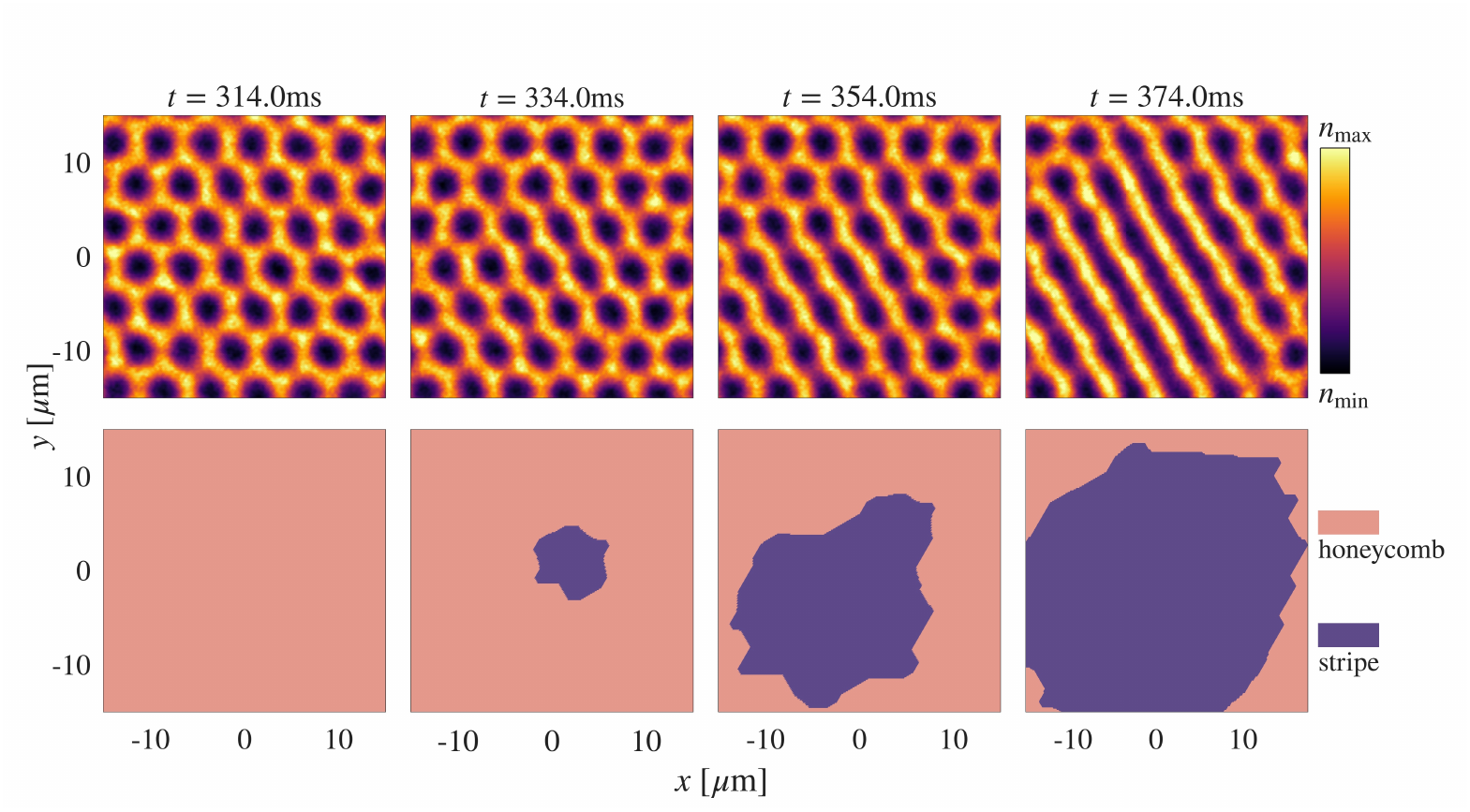}
        \caption{Bubble formation. Columns left to right show the formation of a single bubble using the same parameters as in Fig.~\ref{fig:Bubbles}, zoomed-in. The top row shows the real-space density profile, while the identified phases using the anisotropy are shown in the lower row.}
        \label{fig:BubbleZoom}
    \end{figure*}
    
    In our current study, we do not see evidence for critical bubble sizes, but rather that the stripe phase forms on the order of the system unit cell. In Fig.~\ref{fig:BubbleZoom}, we show the same initial bubble as shown in Fig.~\ref{fig:Bubbles}, just before and after nucleation. In this case, bubble nucleation appears to occur from a rearrangement of the superfluid such that honeycomb superfluid `dimples' elongate and merge to form a short valley, or, equivalently, the high superfluid rails straighten to form stripes. While both endpoints (i.e. $t=314.0\,\mathrm{ms}$ and $t=374.0\,\mathrm{ms}$) are relatively easy to visually classify, the intermediate time slices are more subtle, since distorted remnants of the honeycomb remain as low-density connections. The binary identification in the bottom row of Fig.~\ref{fig:BubbleZoom} in the intermediate cases indicates the general difficulty with classification of the different morphologies in an unambiguous way. The anisotropy parameter we use to identify the phases point towards the fact that the critical bubble is therefore only on the order of one or two lattice sites. Although our system is continuous, the underlying lattice therefore provides two features of bubble formation: (i) a fundamental size limit to the identification of different morphological supersolid phases and (ii) a preferred stripe direction after nucleation. The continuous evolution of the honeycomb lattice via distortion also imposes a challenge when it comes to the identification of an interface, particularly in the context of critical bubbles. A potentially interesting direction would be to further examine the stability of the honeycomb phase via lattice distortions, and how excitations of the system (e.g. lattice phonons) lead to critical distortions seeding nucleation.

	%==============================================================================
    \subsection{Bubble front speed}

   % In the following, we quantitatively analyze the growth rates of supersolid stripe bubbles, focusing on the speed at which the bubble surface expands. The rate at which bubbles spread after formation is typically set by causality: in cosmology this corresponds to the speed of light \cite{Coleman1977FateI}, meanwhile in a superfluid it is the speed of sound \cite{Fialko2015Fate}. In the case of a supersolid, the natural question becomes: which speed of sound? Supersolids possess a superfluid Goldstone (sound) branch associated with the breaking of a global $\mathrm{U}(1)$ symmetry, as well as additional crystal Goldstone branches associated with each broken translational symmetry. Thus, the honeycomb phase has two crystal phonon branches, while the stripe phase only has one. Example spectra within the first Brillouin zone for the stripe and honeycomb states are shown in App.~\ref{app:BogTheory}.

   In the following, we quantitatively analyze the growth rates of supersolid stripe bubbles, focusing on the speed at which the bubble surface expands. The rate at which bubbles spread after formation is typically set by causality: in cosmology this corresponds to the speed of light \cite{Coleman1977FateI}, meanwhile in a superfluid it is the speed of sound \cite{Fialko2015Fate,Canaletti2024Seeding,Jenkins2024Analog}, identified by mapping the superfluid action to a Lorentz-invariant form. In our case, we make no assumptions or connection to Lorentz-invariance, however we will proceed by studying how fast bubbles grow in our system, and compare this speed against the range of speeds of sound in the supersolid.
    
	%==================================================
    \begin{figure}
        \centering
        \includegraphics[width=1.0\linewidth]{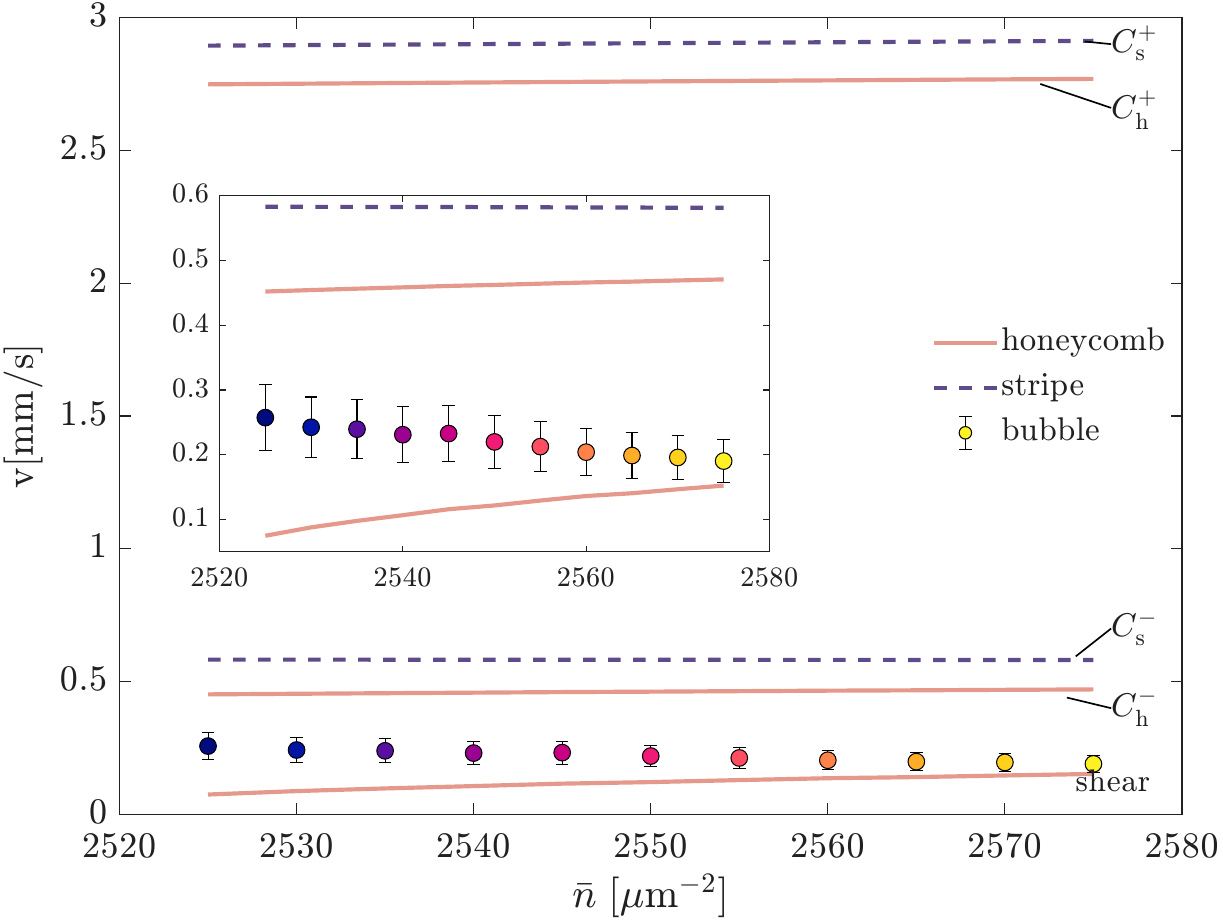}
        \caption{Speeds of sound and bubble growth velocity at $a_\mathrm{s}=102.25\,a_0$ for the same density range as Fig.~\ref{fig:BubbleAreas}. Bubble speeds are estimated using Eq.~\eqref{eq:Bubble_v} at a time $t_\star$ where we define $\mathcal{A}(t_\star)=0.1$, with the error bars corresponding to the upper and lower bounds of the inequality. These front speeds are compared against the speeds of sound in the honeycomb phase (red, solid lines) from slowest to fastest: shear, $C_{\mathrm{h}}^-$ (also called superfluid, or second sound), and $C_{\mathrm{h}}^+$ (crystal, or first sound). The stripe phase has only two speeds of sound (blue, dashed): $C_{\mathrm{s}}^-$ and $C_{\mathrm{s}}^+$, plotted specifically for modes perpendicular to the stripes (i.e. along $k_y$ in Fig.~\ref{fig:BdGSpectrum}). Near the stable regime, the estimated bubble growth speeds approach the shear sound velocity of the honeycomb phase, but deviates as this mode softens towards instability. The inset is the same data, zoomed-in.
        }
        \label{fig:SoundSpeeds}
    \end{figure}
	%==================================================
    
    Numerically, it is not simple to extract the front speed directly, however once regions are identified we do have access to each individual bubble perimeter and the respective area. Assuming roughly circular bubble growth for a total area of $\mathcal{A}$, then we can estimate an upper and a lower bound on the speed of bubble formation,
    \begin{equation}
        \frac{1}{\sum_iP_i}\frac{\mathrm{d}\mathcal{A}}{\mathrm{d}t}\lesssim \mathrm{v}\lesssim \frac{1}{\sum_i2\sqrt{\pi A_i}} \frac{\mathrm{d}\mathcal{A}}{\mathrm{d}t}\;,
        \label{eq:Bubble_v}
    \end{equation}
    where $\mathrm{v}$ is the bubble front growth speed, $P_i$ is the perimeter of the $i$th bubble with area $A_i$, with the sum being over all bubbles in the simulation. For a circular bubble of radius $r_i$, one has $A_i=\pi r_i^2$, giving $\mathrm{d}A_i/\mathrm{d}t=P_i \mathrm{v}$ for fixed bubble growth speed $\mathrm{v}=\mathrm{d}r_i/\mathrm{d}t$. Rearranging and summing over all bubbles ($\sum_iA_i=\mathcal{A}$) gives the exact speed of growth for a circular bubble: $({\sum_iP_i})^{-1}{\mathrm{d}\mathcal{A}}/{\mathrm{d}t}=\mathrm{v}_{\mathrm{circ}}$. If the bubble is noncircular, $P_i$ is larger than $2\pi r_i$ at fixed bubble area $A_i$, and therefore it will decrease the estimate of $\mathrm{v}$ relative to a perfectly circular bubble, providing the lower bound. For the upper bound, we consider first again a circular bubble at fixed perimeter $P_i=2\sqrt{\pi A_i}$, and we use the fact that perfectly round bubbles maximize the area at fixed perimeter, meanwhile other shapes will diminish the denominator on the right, giving an upper bound on the estimate of the front speed. Approaching the analysis by considering all possible independent bubbles, we do not need to exclude multi-nucleation events. 

    In Fig.~\ref{fig:SoundSpeeds} we show the extracted bubble front speed $\mathrm{v}$, with the data points marking the average of the upper and lower bounds from \eqref{eq:Bubble_v} as error bars, for the same data and thus the densities shown in Fig.~\ref{fig:BubbleAreas}. The bubble speeds are numerically extracted at the time $t_\star$ at which $10\%$ of the system has decayed to the stripe phase [i.e.~we define $\mathcal{A}(t_\star)=0.1$], so that we have reasonably sized bubbles, but still well before any finite-size effects disrupt the bubble growth. The quantity $\mathrm{d}\mathcal{A}/\mathrm{d}t$ in Eq.~\eqref{eq:Bubble_v} is extracted by taking a numerical gradient of the mean curve $\mathcal{A}(t)$ at $t_\star$. We see similar values of $\mathrm{v}$ (identical within error bars) and a qualitatively similar general trend, by choosing to extract the speeds at different stages of bubble formation up to approximately $\mathcal{A}=0.3$ (see App.~\ref{app:Robustness}). Beyond this point, we estimate that finite-size effects in our simulations and bubble merging events begin to be highly relevant.
    
    We compare the extracted front speeds against each of the speeds of sound for both the stripe and honeycomb configurations, determined by Bogoliubov-de Gennes calculations, for $a_\mathrm{s}=102.25\,a_0$. For details on how these calculations are performed, as well as example supersolid spectra, see App.~\ref{app:BogTheory} (see also Refs.~\cite{Watanabe2012Spontaneous,Kunimi2012Meanfield,Poli2024Excitations,SenarathYapa2025Anomalous,Blakie2025Dirac,Cook2026Excitations,Poli2026Sound}). Breaking of continuous symmetries generally result in the emergence of an associated Goldstone (sound) mode \cite{Goldstone1961Field,Goldstone1962Broken}. Supersolids therefore possess a phonon branch with the breaking of a global $\mathrm{U}(1)$ symmetry, as well as additional branches associated with each broken translational symmetry. The honeycomb phase has three total phonon branches, while the stripe phase only has two, since the stripe state is essentially a 1D supersolid along its modulated direction, and a uniform superfluid parallel to the stripes. Since there is no dipolar tilt (and hence no preferential direction), there is also no restoring force to a long-wavelength transverse shear of the stripes. The stripe phase therefore only supports a single hard phonon branch parallel to the stripe orientation \cite{Cook2026Excitations,Poli2026Sound}.
    
    In the metastable honeycomb state, we identify three phonons: a transverse shear branch, and two longitudinal modes that we designate by $C_{\mathrm{h}}^-$ and $C_{\mathrm{h}}^+$. As can be seen at the bottom of Fig.~\ref{fig:SoundSpeeds}, all three honeycomb modes are stable (non imaginary) in the metastable regime. However, the shear mode softens and drops to 0 as one approaches the instability boundary, i.e. towards lower densities. The point where the shear speed drops to zero ultimately marks the edge of the region where the metastable honeycomb state can be supported while stripes are the true ground state. For a more complete discussion of the shear instability, see Ref.~\cite{Blakie2025Dirac}. In the honeycomb phase, the remaining modes can largely be differentiated as based on their crystal or superfluid character. The upper branch $C_{\mathrm{h}}^+$ is sometimes called `density' branch, or first sound \cite{Poli2024Excitations,Blakie2025Dirac,SenarathYapa2025Anomalous} (see also Ref.~\cite{Kirkby2024Excitations}), corresponding to compression and expansion of the lattice. Meanwhile, the lower branch $C_{\mathrm{h}}^-$ is sometimes called the superfluid/'phase' branch, or second sound. These modes are dominated by flow of superfluid between lattice sites.

    For the stripe phase, we similarly designate $C_{\mathrm{s}}^-$ and $C_{\mathrm{s}}^+$. Considering modes parallel to the stripes, the system is a uniform superfluid and therefore only supports a single longitudinal phonon, $C_{\mathrm{s}}^+$. Here, although shear-type excitations are possible, there is no associated phonon ($C_{\mathrm{s}}^-\to 0$) and the dispersion for this branch becomes free-particle in nature, i.e. $\propto k^2$ as $k\to0$ \cite{Kunimi2012Meanfield,Poli2026Sound,Cook2026Excitations}. In the stripe phase, there is evidence for strong hybridization in the mode character, particularly in between the two stripe axis directions, and no mode has a purely superfluid or crystal character \cite{Cook2026Excitations,Poli2026Sound}. For the purposes of our study, however, we only concern ourselves with the various speeds of sound rather than mode character.
    
    Fig.~\ref{fig:SoundSpeeds} shows that the bubble growth rate is near to the lowest speed of sound in the system, namely the shear mode in the honeycomb phase. However, in contrast with the softening of this mode (inset) towards the instability, the bubble front speed is found to grow slightly with decreasing density. This behavior is not yet understood. However, for certain classical nucleation theories, domain growth of metastable phases occurs at a speed proportional to the energy difference in the two states \cite{Grossmann1991Kinetic,Clerc2008Transversal,Haudin2009Driven}, which in our system increases as the mean density is lowered. Our model is closely connected with the classical Hohenberg-Halperin Model A, corresponding to the $\gamma\gg 1$ limit of the SPeGPE \cite{Hohenberg1977Theory,Valls1990Nucleation,Groszek2021Crossover}. Although in our system the velocity increases as the energy difference between the  ground and metastable states increases, we are unable to verify any further physical relationship between these quantities. This may be interesting to explore in future works, particularly in combination with experiment where the dissipation parameter $\gamma$ can be benchmarked.
    %Meanwhile, in such theories, another relevant consideration is the modification of $\mathrm{v}$ due to bubble surface tension. In the stripe-honeycomb system we consider here, it is not immediately clear to us how the bubble surface can be identified, or whether surface tension can be calculated using the tools we have developed.
    %Close to the shear instability, the transverse sound speed falls towards zero, while the bubble front speed apparently grows slightly (inset). 
    
    It should be noted that in general the speeds of sound are not unique, since there is a preferred direction set by the crystal orientation. In the honeycomb, the difference in phonon velocities between each direction is relatively marginal. In the stripe state, the upper speed of sound $C_{\mathrm{s}}^+$ does not change drastically with direction, however the lower branch $C_{\mathrm{s}}^-$ varies strongly with direction (see App.~\ref{app:BogTheory}), so Fig.~\ref{fig:SoundSpeeds} shows only the upper limit of $C_{\mathrm{s}}^-$ for clarity.
    
    In the honeycomb phase, the dispersion of shear waves can also yield an \textit{anomalous} dispersion \cite{Maris1973Hydrodynamics,Dynes1975Measurement,SenarathYapa2025Anomalous}: an increase in phononic group velocity at higher $k$, leading to supersonic excitations at higher momenta, and allowing for phonon (Beliaev) decay \cite{Maris1977Phonon,Pitaevskii16}. This effect is also seen in optics, where the refractive index increases with longer wavelengths, flipping the rainbow order, and is in contrast with \textit{normal} dispersion, where the phonon branch bends downwards with higher wavenumber. We find that in the metastable honeycomb phase, the anomalous nature of the shear branch becomes more pronounced as the instability is approached.
    
	%==============================================================================
	%==============================================================================
    \section{Decay rates}
    \label{sec:Rates}

    In this section we compare the numerically extracted false-vacuum decay rate against a theoretical model based on deriving an effective action and solving for the bounce solution interpolating between the two morphologically distinct supersolid orders. 

	%==============================================================================
    \subsection{Numerical decay rate}
\begin{figure}
        \centering
        \includegraphics[width=1.0\linewidth]{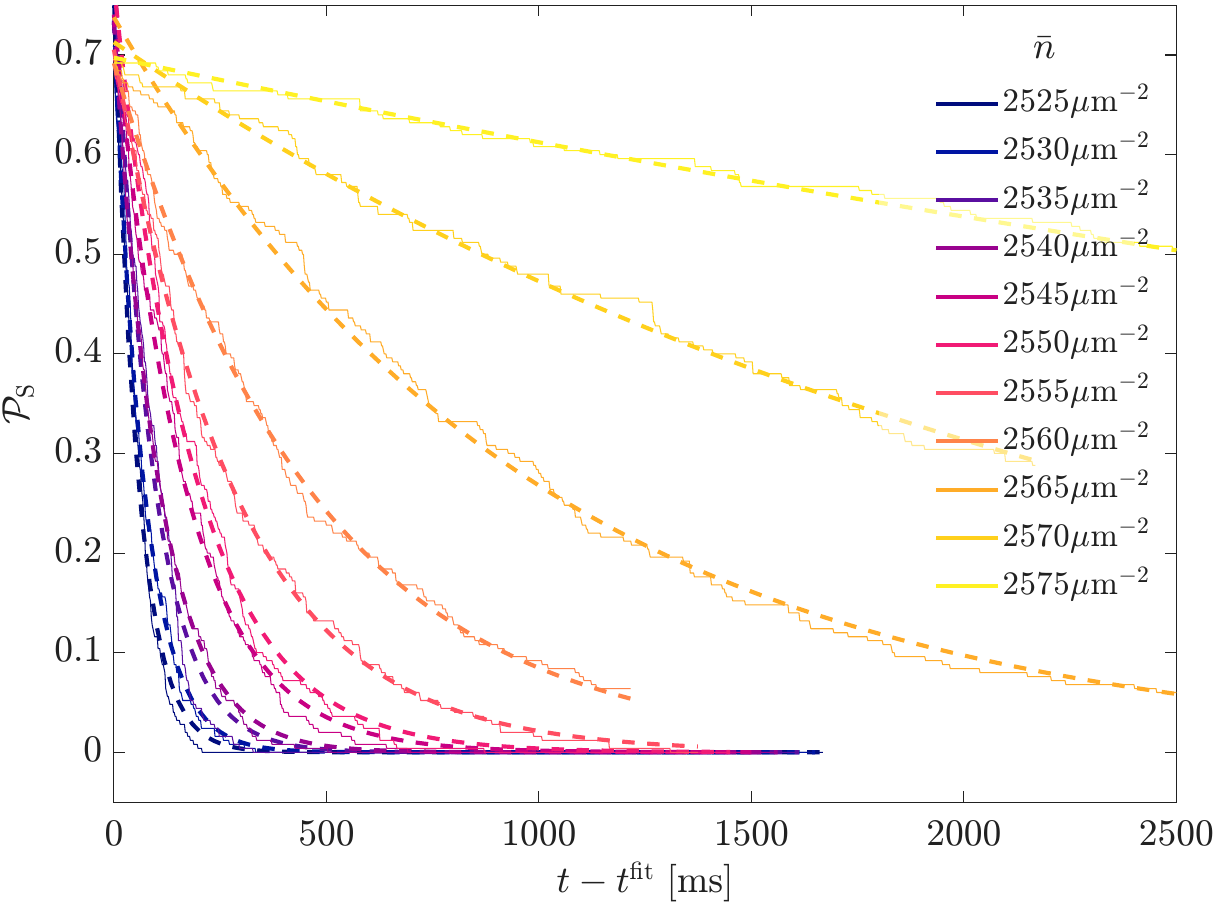}
        \caption{Survival probability $\mathcal{P}_\mathrm{S}(t)$ of the false vacuum (solid lines), for $a_\mathrm{s}=102.25\,a_0$ and the same 250 simulations as in Fig.~\ref{fig:BubbleAreas}. Each data set for $\mathcal{P}_\mathrm{S}(t)$ is shifted by the time $t^{\mathrm{fit}}$ at which the numerical data drops below 0.7, where we begin the respective fit to an exponential function (dashed lines).
        }
        \label{fig:Survival}
    \end{figure}

    In order to extract a decay rate from our SPeGPE simulations, we fit the the survival probability of the metastable honeycomb to a decaying exponential, $\mathcal{P}_\mathrm{S}(t)\propto\,\mathrm{e}^{-\Gamma t}$. In general, this exponential form is not guaranteed from $t=0$, but occurs after some onset time \cite{Billam2019SimulatingSeeded,Billam2022Falsevacuum,Jenkins2024Analog,Zenesini2024False,Sivasankar2025Temperature} so we fit at times $t>t^{\mathrm{fit}}$, where the numerically-extracted survival probability obeys $\mathcal{P}_\mathrm{S}(t)\leq 0.7$. This typical delay in the onset of exponential may be due to the fact that there is a delay before the collective Bogoliubov modes populate after $t=0$, since the SPeGPE includes Gaussian noise suddenly on the unperturbed background; however it may also be related to deviations in exponential tunneling at short times \cite{Wilkinson1997Experimental}.\footnote{Note that although they are similar in magnitude, $t^{\mathrm{fit}}$ is a different quantity than $t_n$ used in Fig.~\ref{fig:BubbleAreas}. The latter is the mean nucleation time and was used simply to overlap and compare nucleation area curves.} Using $t^{\mathrm{fit}}$, the rate $\Gamma$ can then be compared against the theory extracted from an instanton bounce solution. In Fig.~\ref{fig:Survival}, we plot the survival probability of the metastable honeycomb state as a function of time, which we extract as the relative number of trials (i.e. out of the 250 total trials for each parameter set) that do not have more than 5\% identified stripe area. Each full decay curve is shifted by $t^{\mathrm{fit}}$ to the fitting region for easier visibility. We perform simulations over a range of total times, depending on the density. Our longest simulation time is a total of $8\,\mathrm{s}$, by which all except the highest densities have largely decayed. The higher densities seem to require extremely long times to decay, which correspond to unfeasibly long hold times for a dipolar supersolid in an experimental setting and are therefore excluded from our simulations here. We also note here that variations in $\gamma$ tend to set overall timescales in the SPeGPE \cite{Bland2022TwoDimensional}, while they do not affect scaling laws \cite{Kasamatsu2003Nonlinear,Damski2010Soliton,Patra2026Kibble}. We therefore aim to establish qualitative trends with respect to the scaling of rates in our system, with the true physical rates being subject to a direct comparison with experiment.

	%==================================================
	%==================================================

    % \subsection{Effective action and bounce}
    % \label{subsec:Bounce}

	%==============================================================================
    \subsection{Effective action model}
    \label{subsec:effectiveaction}
    In order to develop a model of bubble formation rates, we begin by introducing a parametrization in terms of the real parameters $c_{i}$ \cite{vanZyl2015Density,Lima2025Supersolid},
    \begin{align}
        \psi(x,y)=\sqrt{\bar{n}}\Bigg[
        &c_0
        -c_1\cos\left(\frac{2ky}{\sqrt{3}}\right)
        -2c_2\cos\left(\frac{ky}{\sqrt{3}}\right)\cos(kx)\Bigg]\,,
        \label{eq:ansatz}
    \end{align}
    where $k=2\pi/L_{\mathrm{uc}}$, $\bar{n}$ is the mean density, and $L_{\mathrm{uc}}$ is the side length of a single rhombus unit cell, and $c_0=[1-c_1^2/2-c_2^2]^{1/2}$ ensures normalization for the number of particles within a unit cell such that \eqref{eq:meandensity} is fulfilled.
    %$\frac{2}{\sqrt{3}L_{\mathrm uc}^2}\int_{\mathrm{uc}}\mathrm{d}\mathbf{x}|\psi|^2=\bar{n}$. 

	%==================================================
    \begin{figure}
        \centering
        \includegraphics[width=0.98\linewidth]{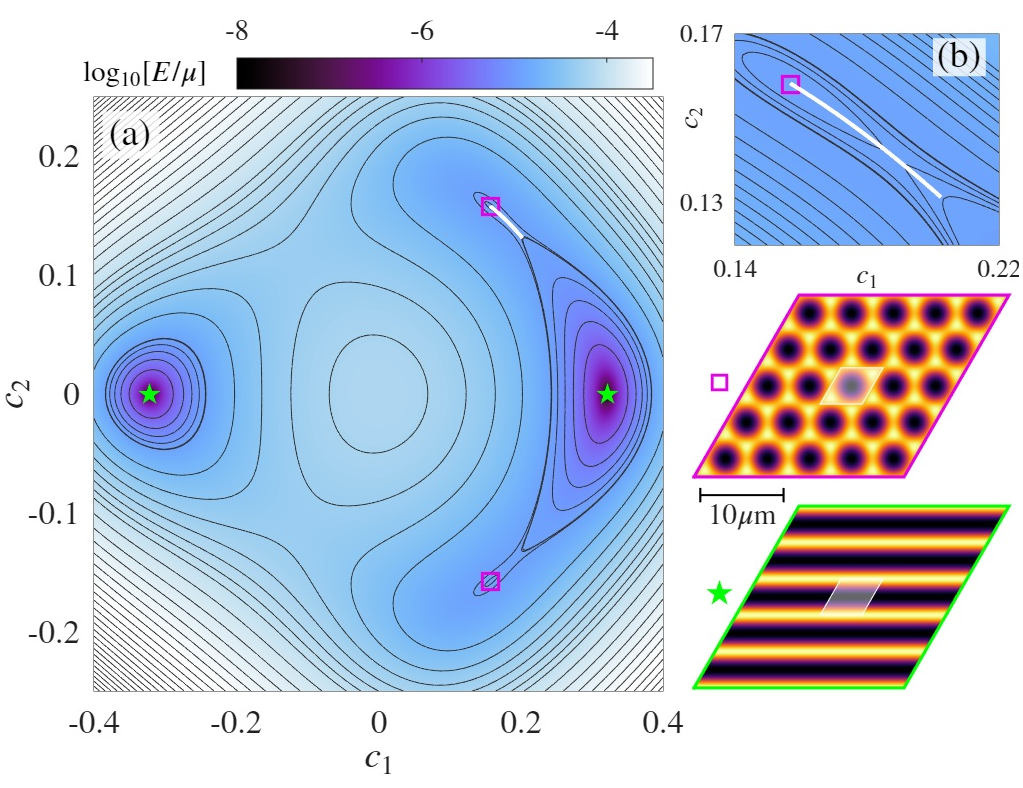}
        \caption{Equipotential contour plot for our 2D ansatz in $\mathbf{c}$-space, for $a_\mathrm{s}=102.25\,a_0$ and $\bar{n}=2575\,\mu \mathrm{m}^{-2}$, corresponding to a chemical potential of $\mu=66.4\,\hbar\omega_z$. Panel (a) shows a relatively large region, while panel (b) focuses on the false-vacuum region and the bounce path, shown as a white line in both panels, and which starts at the false vacuum, ends at the equipotential line $E=E_{\rho}(\mathbf{c}_{\mathrm{F}})$, and represents a stationary point of the action, $\delta B[\bar{\mathbf{c}}]=0$. The remaining panels show the supersolid density $|\psi|^2$ profile given by Eq.~\eqref{eq:ansatz} at the true $(\bigstar)$ and false $(\square)$ vacuua. In each, the rhombus primitive unit cell is highlighted.}
        \label{fig:Countours}
    \end{figure}
	
    %\footnote{Recall this is not a quasi-2D theory, but rather reduced 3D via a variational parameter \cite{Ripley2023Twodimensional}.}
    %^doesn't work properly in figure captions
    %==================================================

    The ansatz~\eqref{eq:ansatz} projects the system to a two-dimensional energy landscape defined by the Gross-Pitaevskii energy functional $E_{\mathrm{eGP}}[\psi]\to E_{\mathrm{eGP}}(\mathbf{c})$, shown in Fig.~\ref{fig:Countours}. The amplitudes $\mathbf{c}=(c_1,c_2)$ connect different configurations of the superfluid modulation. The choice $\mathbf{c}=(0,0)$ results in a uniform condensate, corresponding to an unstable maximum in the energy landscape. Meanwhile, the points $\mathbf{c}=(\pm c_\text{S},0)$ are global minima corresponding to the stripe phase. Finally, the two points $(c_\text{H},\pm c_\text{H})$ give the field in the honeycomb phase, which for certain values of the density and scattering length are local minima and thus correspond to false vacua. In each of these cases, $c_\text{S}$ and $c_\text{H}$ take values which are unknown upfront and must be determined by minimization of the energy functional $E_{\mathrm{eGP}}(\mathbf{c})$ (explicit expressions are given in App.~\ref{app:ParamSpace}), for a given mean density and scattering length. In turn, the minima determine the superfluid fraction of the underlying state. In Fig.~\ref{fig:Countours} (b), one sees that there is indeed a slight energy barrier that the bounce path must cross to move towards the true vacuum.\footnote{The absolute height of this barrier in our 2D theory cannot be reliably compared against the SPeGPE temperature, since we find that, e.g., at $\bar{n}=2575\,\mu\mathrm{m}^{-2}$, the unit-cell barrier height is $T_{\mathrm{barrier}}\approx 0.5\,\mathrm{nK}$. This would imply that FVD would be thermally triggered instantly, but it does not account for the full configuration space of the fields, dissipation via the thermal bath, nor critical bubble sizes. See also \cite{Sivasankar2025Temperature,Billam2022Falsevacuum}, and App.~\ref{app:Temperature}.}

    In Coleman's original theory \cite{Coleman1977FateI}, the vacuum decay rate is related to the \textit{bounce path}, i.e. an evaluation of the action along a path in imaginary time, starting from the false vacuum. The bounce solution responsible for bubble formation can be calculated starting from the corresponding imaginary-time, i.e. Euclidean, effective action in terms of the time-dependent coordinates $\mathbf{c}(\tau)$ (for details of the derivation cf.~App.~\ref{app:BounceDerivation}),
    \begin{equation}
        S[\mathbf{c}]
        =\int\!\mathrm{d}\tau\left[-\frac{1}{2}\sum_{ij}\dot{c}_iM_{ij}\dot{c}_j
        +E_{\rho}(\mathbf{c})\right]\,,
        \label{eq:Seffc}
    \end{equation}
    where the effective Hamiltonian energy $E_{\rho}$ is the sum of kinetic and interaction energies of the density variations only, cf.~\eqref{eq:Edensity}. The resulting action corresponds to motion on a curved manifold equipped with a metric tensor,
    \begin{equation}
        M_{ij}(\mathbf{c})=\frac{\hbar^2}{m}\int\!\mathrm{d}\mathbf{x}\;\rho\nabla\chi_i\cdot\nabla \chi_j\;,
        \label{eq:Metric}
    \end{equation}
    where the functions $\rho(\mathbf{x},\mathbf{c}(\tau))$, and $\chi_i(\mathbf{x},\mathbf{c}(\tau))$, now depend on $\tau$ through $\mathbf{c}$ and are solutions to the imaginary-time continuity equations
    \begin{equation}
        \mathrm{i}\frac{\partial \rho}{\partial c_i}+\frac{\hbar}{m}\nabla\cdot\left[\rho\nabla\chi_i\right]=0\;.
        \label{eq:ChiContinuity}
    \end{equation}
    Note that in this formalism, there is no explicit need to only consider a two-dimensional parameter space ($i=1,2$), but as we shall see, this minimal model is sufficient to capture FVD rate behaviour within the part of the phase diagram we consider.
    %The Gross-Pitaevskii energy function $E_{\mathrm{GP}}(\mathbf{c})$, corresponds to the spatial integral of the energy densities using the ansatz Eq.~\eqref{eq:ansatz} and is an analytic function of $\mathbf{c}$.

    In the above, we have defined the (as-of-yet unknown) functions $\chi_i$, corresponding to solutions of Eq.~\eqref{eq:ChiContinuity}, which connect the local phase of the full wavefunction $\psi(\mathbf{x},\tau)=\sqrt{\rho(\mathbf{x},\tau)}\mathrm{e}^{\mathrm{i}\theta(\mathbf{x},\tau)}$ to the imaginary-time-dependence of the density via,
     \begin{equation}
        \theta(\mathbf{x},\tau)=\sum_{i=1}^{\mathcal{N}}\dot{c}_i(\tau)\chi_i(\mathbf{x},\mathbf{c})\;.
        \label{eq:theta_chi}
    \end{equation}
    This assumption comes from the fact that the overall phase $\theta(\mathbf{x},\tau)$ cannot rely only on any instantaneous values of $c_i$ (since it is not unique up to a constant), but nonzero $\dot{c}_i$ indicate a restructuring of the wavefunction itself in (imaginary) time, and therefore a nonuniform phase due to the corresponding superfluid flow. Put differently, if $\dot{c}_i=0$ the wavefunction structure is constant and then at least within our ansatz parameter space, the phase is a constant that we can set arbitrarily to 0. Thus, up to a space-time dependent factor $\chi_i$, the phase should at least be dependent on $\dot{c}_i$, giving Eq.~\eqref{eq:theta_chi}. Note again that while we limit our parameter space to two dimensions, this generalizes to arbitrary basis functions with weights $c_i$.

    According to Coleman's instanton theory, at leading order in $\hbar$, the false-vacuum nucleation rate $\Gamma$, per unit area, is well approximated by,
    \begin{equation}
        \Gamma=A\,B[\bar{\mathbf{c}}]\,\mathrm{e}^{-B[\bar{\mathbf{c}}]/\hbar}\,,\label{eq:GammaBounce}
    \end{equation}
    where $A$ corresponds to a functional determinant that we shall take as a single fit parameter for our decay rate model.
    $B[\bar{\mathbf{c}}]$ is the action \eqref{eq:Seffc} along the bounce path $\bar{\mathbf{c}}(\tau)$ relative to the action at the false vacuum, $\mathbf{c}_F=$ const.,
    \begin{equation}
        B=\int\!\mathrm{d}\tau\left[-\frac{1}{2}\sum_{ij}\dot{\bar{c}}_iM_{ij}(\bar{\mathbf{c}})\dot{\bar{c}}_j+E_{\rho}(\bar{\mathbf{c}})-E_{\rho}(\mathbf{c}_\mathrm{F})\right]\,.
        \label{eq:B1}
    \end{equation}
    The apparent unboundedness due to the negative sign of the kinetic energy in both Eqs.~\eqref{eq:Seffc} and \eqref{eq:B1} is resolved by the fact that currents in imaginary time give imaginary $\chi_i$, which results in an overall positive contribution. Energy conservation along the bounce path allows us to write the bounce action as,
    \begin{equation}
        B=2\int\!\mathrm{d}\mathbf{s}\sqrt{2\left[E_{\rho}(\bar{\mathbf{c}})-E_{\rho}(\mathbf{c}_\mathrm{F})\right]}\;,\label{eq:BounceE}
    \end{equation}
    where $(\mathrm{d}\mathbf{s})^2=-\sum_{ij}\mathrm{d}c_iM_{ij}\mathrm{d}c_j$, and the integral runs along the bounce path, starting at the turning point and ending at the false vacuum. The overall factor of 2 comes from the fact that the bounce action is twice the action along the one-way path. Note that Eq.~\eqref{eq:BounceE} takes the form of a WKB phase picked up along the bounce path.

    The bounce path can be calculated by solving the classical equations of motion associated with the action \eqref{eq:Seffc}, given explicitly in \eqref{eq:Eqddot1}, \eqref{eq:Eqddot2}. In Fig.~\ref{fig:Countours}, we plot the bounce trajectory in $\mathbf{c}$-space as a short white line. This solution corresponds to a trajectory starting at the false vacuum (a local peak in the inverted potential landscape), rolling toward the true vacuum to the turning point lying on the equipotential line $E=E_{\rho}(\mathbf{c}_{\mathrm{F}})$, and rolling back again, with boundary conditions $\dot{\bar{\mathbf{c}}}(\tau\to0)=\dot{\bar{\mathbf{c}}}(\tau\to\infty)=0$ and subject to the metric $M_{ij}$. Due to the symmetry in $\pm c_2$, it does not matter from which false vacuum we begin, since the resulting action will be the same. It is far more likely for the particle to tunnel to the positive $c_1$ well, since this is separated by a much lower energy barrier, and so we only calculate the action along that single path.

	%==================================================
    \begin{figure}
        \centering
        \includegraphics[width=1.0\linewidth]{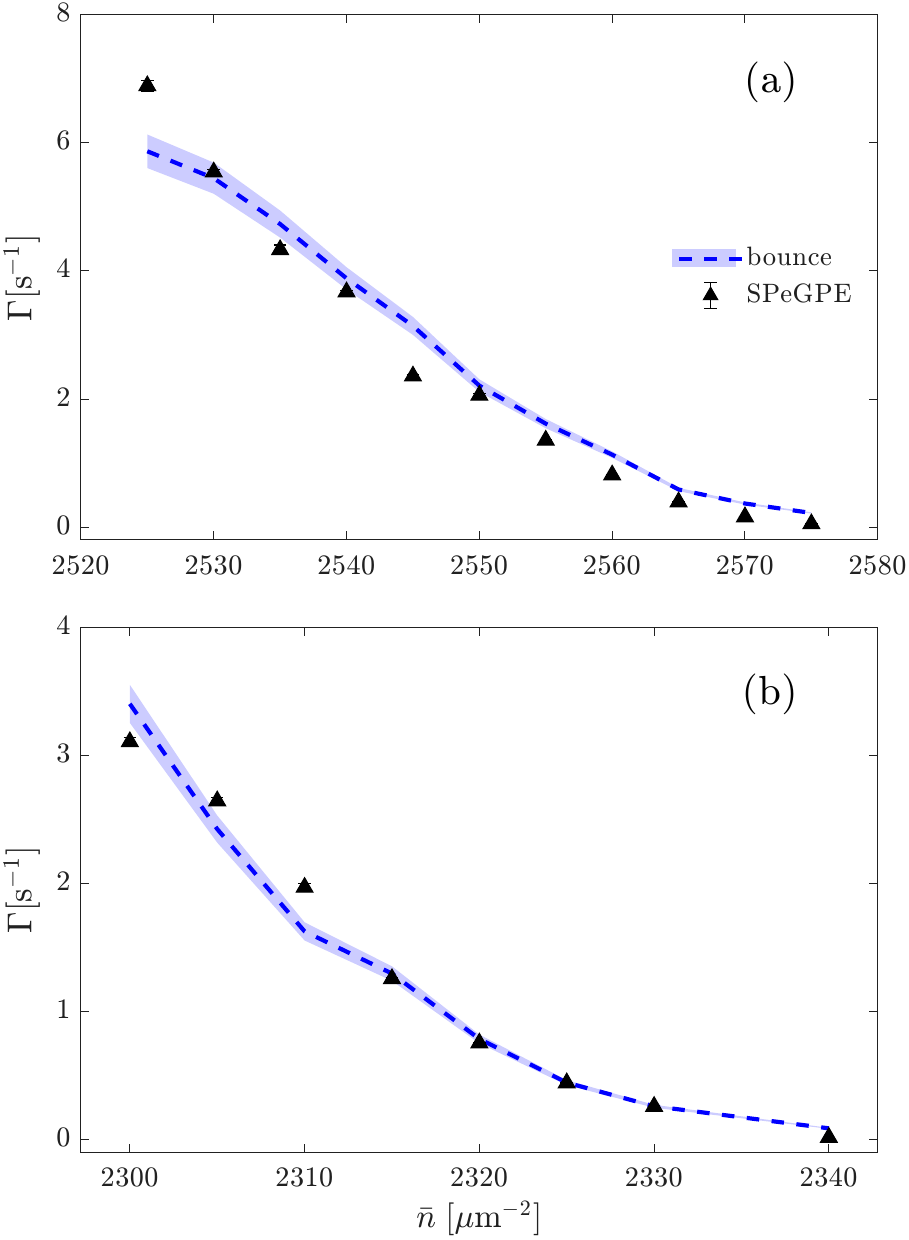}
        \caption{Vacuum decay rate as a function of average density at (a) $102.25\,a_0$  and (b) $102.50\,a_0$. Solid triangles correspond to the numerically extracted decay rate as determined by the survival probability (Fig.~\ref{fig:Survival}) via the SPeGPE simulations. For all parameter sets, we use again 250 trials, with the error bars on the decay rate corresponding to a standard error on the $\Gamma$ fit. The approximate rate defined in Eq.~\eqref{eq:GammaBounce} with an overall amplitude fit $A$ is also shown as a blue dashed line, with the shaded area corresponding to the standard error on the fit.
        }
        \label{fig:DecayRates}
    \end{figure}
	%==================================================

    In Fig.~\ref{fig:DecayRates}, we show the FVD rate versus the average density for two fixed values of the scattering length: (a) $a_\mathrm{s} = 102.25\,a_0$ and (b) $102.50\,a_0$. While these values are close to each other, a small change in the scattering length can result in a large change in superfluid fraction of the stripe phase \cite{Ripley2023Twodimensional}, and have reasonably different density ranges for similar decay rates. We choose these values such that the resulting stripe phase is supersolid. The numerically extracted decay rates, as defined in the previous subsection, are shown as black triangles. Since at higher scattering length the metastable region is smaller, we choose a slightly narrower density window for panel (b). Furthermore, the density contrast is lower for $a_\mathrm{s} = 102.50\,a_0$, such that fluctuations have a greater effect on the density profile. Furthermore, this results in a shallower false-vacuum well (although this is not explicitly calculable without resorting to our ansatz), since low-contrast states are closer to a uniform condensate. Therefore, we reduced the temperature to $5\,\mathrm{nK}$ for these simulations.
    
    We also compare our simulations to the model for the FVD rates in Fig.~\ref{fig:DecayRates}. For the bounce theory, we calculate the action entering Eq.~\eqref{eq:GammaBounce} along the path via Eq.~\eqref{eq:B1} and then fit the overall amplitude $A$ as a single free parameter. In both panels, the minimal two-parameter model appears to perform remarkably well in matching the overall decay trend of the numerically-extracted $\Gamma$ as a function of the density. In App.~\ref{app:Temperature}, we show the effects of higher temperature on the theory, and how our bounce-path approach to the quantum tunneling rate instead fails at higher temperature and so a finite-temperature approach to the instanton theory is instead needed.

    If we tune to lower scattering lengths (not shown), the numerically extracted decay rates obey a similar trend of exponential decrease with density, however our analytic bounce approach fails to capture the qualitative behaviour. We assert that this is due to the fact that at lower scattering lengths, the supersolid configuration with higher density contrast is no longer accurately captured by the lowest Fourier-mode approximation. This is evidenced by the fact that $\mathbf{c}_\mathrm{F}$ no longer exists, even in regimes where the eGPE suggests the presence of a metastable state. It is likely therefore that one could recover the instanton theory by extending the ansatz space to include higher-order modes \cite{Lima2025Supersolid}, noting that the potential $E_\rho$ remains analytic when including second-order Fourier amplitudes. However, as the dimension of the ansatz control space increases, the expressions for the energy become more unwieldy and calculating the bounce paths will be more challenging.

	%==============================================================================
	%==============================================================================
    \section{Conclusions}

    In this work, we have demonstrated FVD of a dipolar supersolid from honeycomb to stripe configurations. The dipolar supersolid provides a novel setting where FVD is directly observable as a structural change in the density profile. Supersolids also offer an intriguing scenario where the analog system has competition between many speeds of sound/causality, even differing between the two phases (i.e. inside vs. outside the bubble), leading to nontrivial bubble-growth front speeds. A deeper understanding of the relationship between bubble growth and the speeds of sound, as well as the possible connection to anomalous dispersion, is worth future examination. A full exploration of stripe phase statistics, including orientation of stripe domains, is also outside the scope of this work but would constitute a natural extension of our analysis.

    Two-dimensional supersolid droplet phases have been realized \cite{Norcia2021Twodimensional,Bland2022TwoDimensional} with the dipole orientation we consider here. The system we study is very large, both in overall size and density, relative to those currently realizable with experimental setups, however we note that this is primarily for numerical and theoretical convenience as a first study of morphological FVD. Comparison directly with experiment will also allow for further benchmarking of the SPeGPE as a dynamical tool for dipolar systems, by establishing physical values for the dissipation parameter $\gamma$ and for the mode projection cutoff. Introducing dipolar tilt allows for greater access to the additional phases by an overall shift in the phase diagram to lower densities \cite{Lima2025Supersolid}, and the experimental observation of supersolid stripe states under dipolar tilt has recently been reported \cite{He2025Observation,Chandrashekara2026Competing}. Dipolar tilt would open the possibility of studying bubble formation speeds in anisotropic conditions, since the speeds of sound in each direction can be tunable by changing relative interaction strengths along a particular axis.
    
    Since the supersolid phase diagram is rich and supports many kinds of transitions, by changing density, scattering length, or dipolar tilt, there are many possible scenarios to realize FVD, of which our setup is only one. The presence of other metastable supersolid phases (see, e.g. \cite{Zhang2024Metastable}) may provide additional platforms to study FVD with unique properties.
    
    In this paper, we also present a relatively simple theory of an instanton restricted to a two-dimensional space based on a cosine-modulated ansatz for the density profile. We find that our model for the instanton bounce fits the trend of the numerical decay rate using a single-parameter amplitude fit at low temperature, indicating that our system is likely dominated by quantum tunneling (or at most, minimally thermally-assisted quantum tunneling) for the temperatures we consider here \cite{Liu2002Quantum} (see Appendix~\ref{app:Temperature}). As shown in the appendices, the phase space describing the field configuation is not limited to only two parameters, although the numerical stability of solutions of the equations of motion for the bounce path in an inverted potential becomes increasingly difficult in higher dimensions. However, the metric seems to have a relatively mild effect on the trajectory through the energy landscape. Thus, a reasonable approximation to the bounce action could possibly be made by neglecting the $M_{ij}$ contribution in higher dimensions and using a simple elastic band method to find the classical path, while still assuming Eq.~\eqref{eq:BounceE} to hold. In this way, extending our model to lower scattering lengths may, in fact, be computationally reasonable.

	%==============================================================================
    \begin{acknowledgements} 
    The authors thank P. B. Blakie, R. Bisset, K. Brown, T. Bland, K. Chandrashekara, J. Gao, C. G\"{o}lzh\"{a}user, F. Kaufmes, J. Kusch, L. Platt, I. Moss, A. Oros, E. Poli, and N. Rasch for helpful discussions and collaboration on related work. They acknowledge support by the Deutsche Forschungsgemeinschaft (DFG, German Research Foundation), through SFB 1225 ISOQUANT (Project-ID 273811115), grant GA677/10-1, and under Germany's Excellence Strategy -- EXC 2181/1 -- 390900948 (the Heidelberg STRUCTURES Excellence Cluster), and by the state of Baden-W{\"u}rttemberg through bwHPC and the DFG through grants INST 35/1503-1 FUGG, INST 35/1134-1 FUGG, INST 35/1597-1 FUGG, and 40/575-1 FUGG (SDS, MLS-WISO, Helix, and JUSTUS2 clusters). W.K. acknowledges the support of the Natural Sciences and Engineering Research Council of Canada (NSERC), [funding reference number PDF-577924-2023]. L.C. acknowledges support by the European Research Council (ERC) under the European Union’s Horizon Europe research and innovation program under grant number 101040688 (project 2DDip). Views and opinions expressed are however those of the authors only and do not necessarily reflect those of the European Union or the European Research Council. Neither the European Union nor the granting authority can be held responsible for them.
    \end{acknowledgements}

	%==============================================================================
    
    \bibliography{DipolarRefs}
    \clearpage

	%==============================================================================
	%==============================================================================
    \begin{appendix}
    \numberwithin{equation}{section}
    % \section{Dimensional reduction}

    % Consider the 3D energy functional for a gas trapped only in the $z$-direction,
    % \begin{align}
    %     E=&\;\int\mathrm{d}\mathbf{r}\Biggl[\frac{\hbar^2}{2m}|\nabla\Psi|^2+\frac{1}{2}m\omega_z^2z^2|\Psi|^2
    %     +\frac{g}{2}|\Psi|^4\nonumber\\&\;+\frac{2}{5}\gamma_{QF}|\Psi|^5+\frac{1}{2}\int\mathrm{d}\mathbf{r}'|\Psi(\mathbf{r})|^2U(\mathbf{x}-\mathbf{x}')|\Psi(\mathbf{r}')|^2\Biggr]
    % \end{align}
    % Inserting our variational ansatz and integrating the $z$-direction gives, the result in Eq.~\eqref{eq:2DEnergy}

	%==============================================================================
    \section{Bogoliubov theory and metastability}
    \label{app:BogTheory}

    The spectrum of the 2D supersolid system may be calculated using Bogoliubov-de Gennes formalism. We make the ansatz,
    \begin{equation}
        \psi(\mathbf{x},t)=\left\{\psi_0(\mathbf{x})+\lambda\left[u(\mathbf{x})\mathrm{e}^{-\mathrm{i}\omega t}-v^*(\mathbf{x})\mathrm{e}^{\mathrm{i}\omega^* t}\right]\right\}\mathrm{e}^{-\mathrm{i}\mu t/\hbar}\,,
    \end{equation}
    where $\psi_0$ corresponds to the ground state, $\lambda$ is some small perturbation strength, and recall $\mathbf{x}=(x,y)$. Inserting this into the extended Gross-Pitaevskii equation, we arrive at the 
    \begin{align}
        \hbar \omega [u(\mathbf{x})+v(\mathbf{x})]=&\;H_0[u(\mathbf{x})-v(\mathbf{x})]+C[u(\mathbf{x})-v(\mathbf{x})]\nonumber\\&+2X[u(\mathbf{x})-v(\mathbf{x})]\,,\\
        \hbar \omega [u(\mathbf{x})-v(\mathbf{x})]=&\;H_0[u(\mathbf{x})+v(\mathbf{x})]+
        F[u(\mathbf{x})+v(\mathbf{x})]\,,
    \end{align}
    with linear operators,
    \begin{align}
        H_0w(\mathbf{x})=&\;\frac{\hbar^2}{2m}\mathcal{F}^{-1}\left[k^2\tilde{w}(\mathbf{k})\right]+U_{\mathrm{ext}}(\mathbf{x})w(\mathbf{x})-\mu w(\mathbf{x})\,,\\
        Fw(\mathbf{x})=&\;\tilde{g}|\psi_0(\mathbf{x})|^2w(\mathbf{x})+\tilde{\gamma}_{Q}|\psi_0(\mathbf{x})|^3w(\mathbf{x})\nonumber\\&+w(\mathbf{x})\mathcal{F}^{-1}\left[\tilde{U}^{2\mathrm{D}}(\mathbf{k})\mathcal{F}\left[|\psi_0(\mathbf{x}')|^2\right]\right]\,,\\
        X[w(\mathbf{x})]=&\;\tilde{g}|\psi_0(\mathbf{x})|^2w(\mathbf{x})+\frac{3}{2}\tilde{\gamma}_{Q}|\psi_0(\mathbf{x})|^3w(\mathbf{x})\nonumber\\&+\psi_0(\mathbf{x})\mathcal{F}^{-1}\left[\tilde{U}^{2\mathrm{D}}(\mathbf{k})\mathcal{F}\left[\psi_0(\mathbf{x}')w(\mathbf{x}')\right]\right] \;.
    \end{align}
    Due to the periodicity of our system, we can make use of Bloch's theorem. We define a crystal momentum $\mathbf{q}$ lying in the first Brillouin zone, such that
    $u(\mathbf{x})=\mathrm{u}(\mathbf{x})\mathrm{e}^{\mathrm{i}\mathbf{q}\cdot\mathbf{x}}$ and $v(\mathbf{x})=\mathrm{v}(\mathbf{x})\mathrm{e}^{\mathrm{i}\mathbf{q}\cdot\mathbf{x}}$. We also define, for convenience,
    \begin{align}
        \mathrm{f}(\mathbf{x})=&\;\mathrm{u}(\mathbf{x})+\mathrm{v}(\mathbf{x})\,,\\
        \mathrm{g}(\mathbf{x})=&\;\mathrm{u}(\mathbf{x})-\mathrm{v}(\mathbf{x})\,.
    \end{align}
    Our Bogoliubov-de Gennes equations become,
    \begin{align}
        \hbar \omega \,\mathrm{f}(\mathbf{x})=&\;H_\mathbf{q}\mathrm{g}(\mathbf{x})+F\mathrm{g}(\mathbf{x})+2X_{\mathbf{q}}[\mathrm{g}(\mathbf{x})]\,,\\
        \hbar \omega \,\mathrm{g}(\mathbf{x})=&\;H_\mathbf{q}\mathrm{f}(\mathbf{x})+F\mathrm{f}(\mathbf{x})\,,
    \end{align}
    with momentum-shifted operators ($F$ remains the same as above),
    \begin{align}
        H_\mathbf{q}\mathrm{w}(\mathbf{x})=&\;\frac{\hbar^2}{2m}\mathcal{F}^{-1}\left[(\mathbf{k}+\mathbf{q})^2\mathrm{\tilde{w}}(\mathbf{k})\right]\nonumber\\&+U_{\mathrm{ext}}(\mathbf{x})\mathrm{w}(\mathbf{x})-\mu\mathrm{w}(\mathbf{x})\,,\\
        X_{\mathbf{q}}[\mathrm{w}(\mathbf{x})]=&\;\tilde{g}|\psi(\mathbf{x})|^2\mathrm{w}(\mathbf{x})+\frac{3}{2}\tilde{\gamma}_{QF}|\psi(\mathbf{x})|^3\mathrm{w}(\mathbf{x})
        \nonumber\\&+\psi(\mathbf{x})\mathcal{F}^{-1}\left[\tilde{U}^{2\mathrm{D}}(\mathbf{k}+\mathbf{q})\mathcal{F}\left[\psi(\mathbf{x}')\mathrm{w}(\mathbf{x}')\right]\right]\,,
    \end{align}
    and thus,
    \begin{align}
        (\hbar \omega)^2 \mathrm{g}(\mathbf{x})=&\;\left(H_\mathbf{q}+F\right)\left(H_\mathbf{q}+F+2X_{\mathbf{q}}\right)\mathrm{g}(\mathbf{x})\,,\label{eq:BdGg2}\\
        (\hbar \omega)^2 \mathrm{f}(\mathbf{x})=&\;\left(H_\mathbf{q}+F+2X_{\mathbf{q}}\right)\left(H_\mathbf{q}+F\right)\mathrm{f}(\mathbf{x})\,\label{eq:BdGf2}.
    \end{align}

	%==================================================
    \begin{figure}
        \centering
        \includegraphics[width=0.95\linewidth]{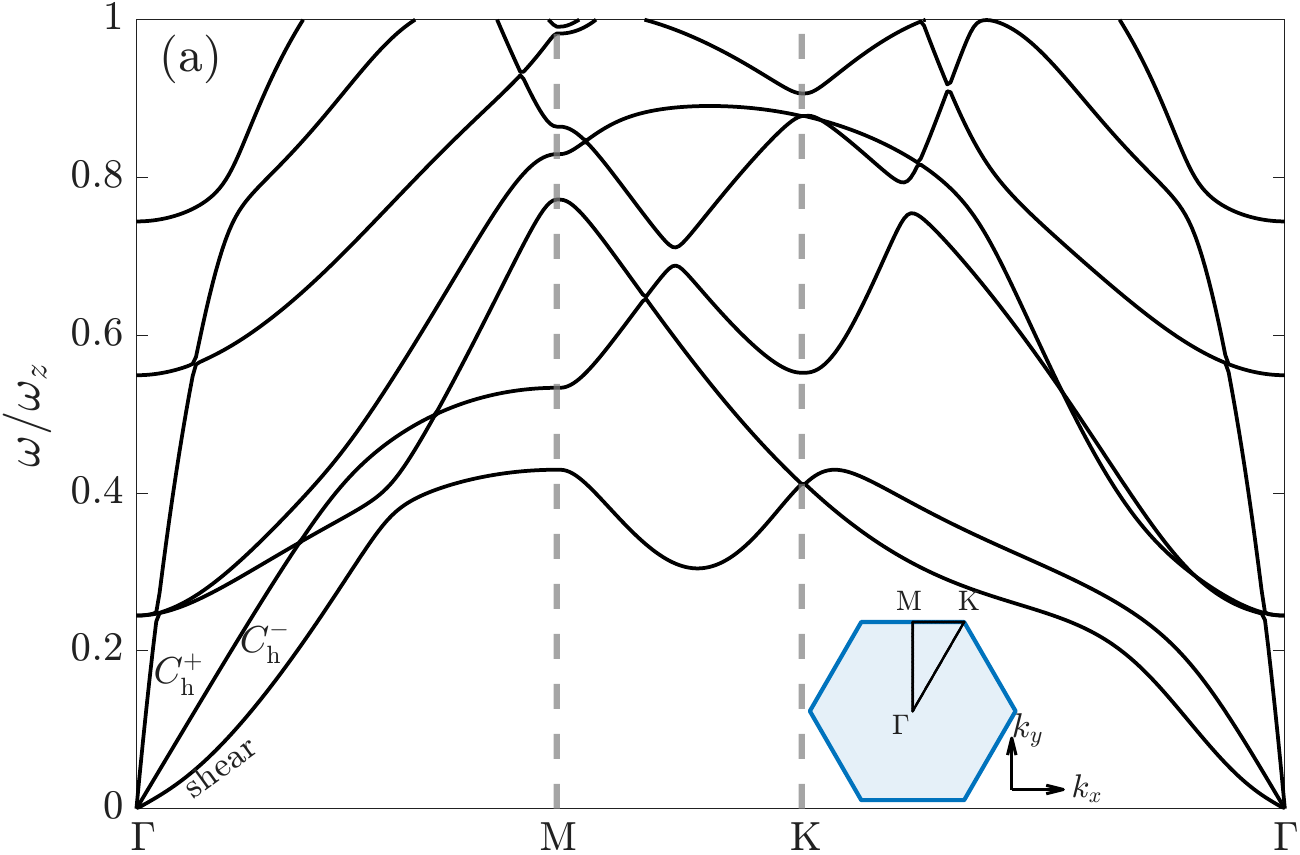}
        \includegraphics[width=0.95\linewidth]{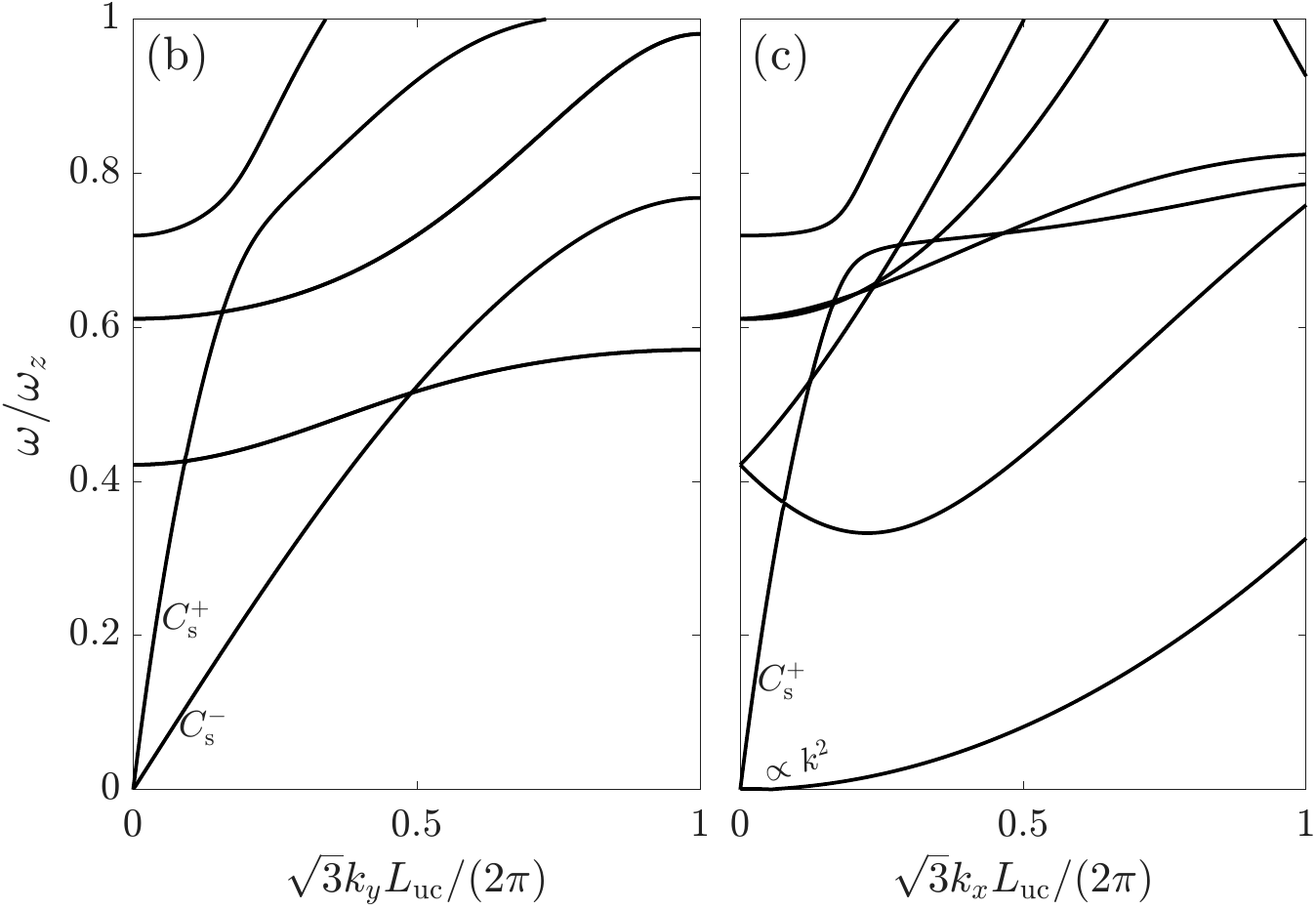}
        \caption{Two-dimensional BdG spectrum for $a_\mathrm{s}=102.25\,a_0$ and $n=2575\,\mu\mathrm{m}^{-2}$. For the honeycomb (metastable) state in (a), we show the first Brillouin zone, along the quasimomentum path shown in the schematic inset. For the stripe (ground) state, for stripes oriented parallel to the $x$-axis, we show excitations (b) perpendicular to the stripes and (c) along the stripe direction. In the stripe phase, sound becomes highly anisotropic. the system supports two sound modes perpendicular to the stripes, $C_{\mathrm{s}}^{+}$ and $C_{\mathrm{s}}^{-}$. Parallel to the stripes, the system only has a single longitudinal sound mode $C_{\mathrm{s}}^{+}$, while $C_{\mathrm{s}}^{-}\to 0$ leading to a shear branch of free-particle nature. At intermediate directions, modes hybridize strongly in character \cite{Cook2026Excitations,Poli2026Sound}.}
        \label{fig:BdGSpectrum}
    \end{figure}
	%==================================================
    
    In Fig.~\ref{fig:BdGSpectrum}, we show an example of the in-plane BdG spectrum, for a quasimomentum path within the first Brillouin zone for both the honeycomb and stripe states. In the honeycomb state, even though it is metastable, three distinct Goldstone modes are clearly visible. These modes are (from slowest to fastest): transverse shear, and two longitudinal modes $C_{\mathrm{h}}^-$, and $C_{\mathrm{s}}^+$. While all three modes remain at least somewhat rigid in the metastable regime, it is the slowest transverse shear mode that develops a phonon instability deeper into the stripe phase, destroying the crystal \cite{Blakie2025Dirac}. In the stripe state, panel (b) shows that there are two stable phonons perpendicular to the stripes, while in (c) one can see that parallel to the stripes only one phonon survives. The branch $\propto k^2$ corresponds to dispersive shear waves along the stripes, and is a manifestation of the stripes' general indifference to bending. When dipoles are tilted into the plane~\cite{Cook2026Excitations}, this branch develops into a phonon.
    
%==============================================================================
    \section{Tracking bubble regions}
    \label{app:RegionIdent}
    
    In contrast with false-vacuum decay in a spin-1 gas, where each region is characterized by a relative phase variable, the different orders considered here are identified by morphologically different field configurations. Tracking regions of different supersolid order can therefore be a numerical challenge. However, the different orders can be distinguished with a reasonable amount of confidence via the local structure tensor, defined as
    \begin{equation}
        T(n)=\begin{pmatrix}
            \left(\frac{\partial n}{\partial x}\right)^2&\frac{\partial n}{\partial x}\frac{\partial n}{\partial y}\\
            \frac{\partial n}{\partial x}\frac{\partial n}{\partial y}&\left(\frac{\partial n}{\partial y}\right)^2
        \end{pmatrix}\label{eq:StructureTensor}\;,
    \end{equation}
    where the eigenvalues $\lambda_1^T$, $\lambda_2^T$ define the normalized anisotropy,
    \begin{equation}
        \Xi=\frac{\lambda_1^T-\lambda_2^T}{\lambda_1^T+\lambda_2^T}\;,
        \label{eq:Anisotropy}
    \end{equation}
    which can reliably distinguish between the two supersolid orders. In particular, a regular stripe phase has a relatively high local anisotropy ($\Xi\approx 0.8-0.9$), while a honeycomb state typically has rather low anisotropy ($\Xi\approx 0.1-0.2$). The elements of $T$ are averaged over small regions of the full crystal, in our case we select the ground-state unit cell size. In order to binarize the output, we select $\Xi=0.8$, as our local cutoff. While precise identification of nucleation can become slightly sensitive to this cutoff, reasonable variations do not affect nucleation times beyond a few milliseconds, with the bubble growth rate being identical at a slightly different threshold. We estimate that this does not affect the main scaling results of our work, since decay rates between different parameter regimes range over thousands of milliseconds.
    
    This method of identification is also sensitive to sudden deviations in the stripe orientation (e.g.~dislocations). An additional filter is applied to identify small regions ($\approx L_{\mathrm{uc}}^2$) of low $\Xi$ surrounded entirely by detected stripe order and assigns their identification as stripes. Large regions of dislocations ignored by this filter are the reason not all curves saturate at $\mathcal{A}=1$, however this long-time behaviour is not the main focus of the manuscript.

    \begin{figure}
        \centering
        \includegraphics[width=\linewidth]{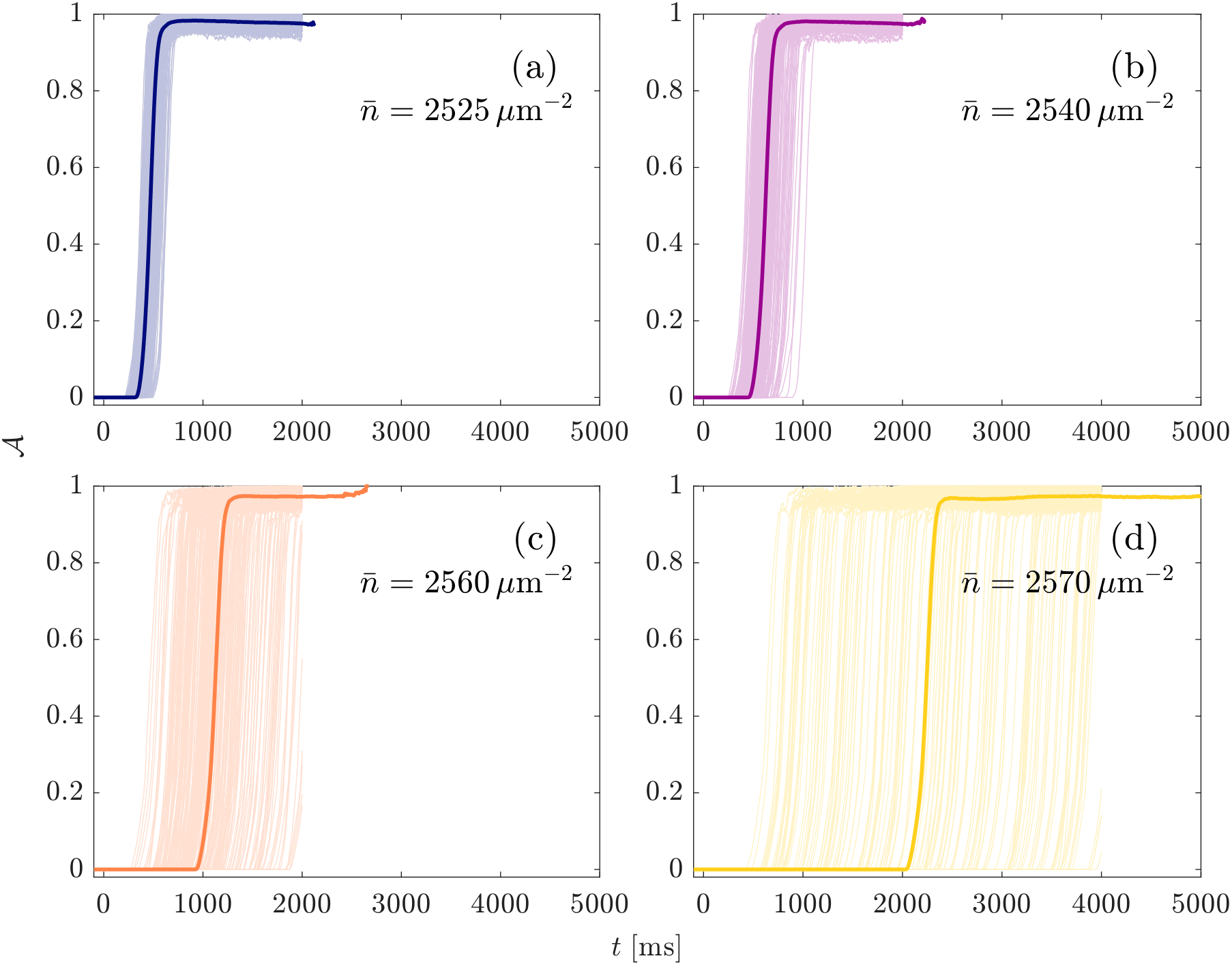}
        \caption{Unshifted raw data for the relative stripe (true vacuum) area versus time for selected densities. Every one of 250 trials for each density is plotted as a semi-transparent line. In order to construct Fig.~\ref{fig:BubbleAreas} of the main text, all of the trial curves are shifted to the same time (selected to be the mean decay time, $\bar{t}_n$) and the mean curve is calculated from the resulting spread. These mean curves (bold, non-transparent) are also shown here in each panel, and are the same as in Fig.~\ref{fig:BubbleAreas}. Note that the choice of where to shift the curves does not affect the later results, and is purely visual.}
        \label{fig:Areas_Unshifted}
    \end{figure}
    
    %==============================================================================
    \section{Stability of decay rates and front speeds}
    \label{app:Robustness}
    In this appendix, we briefly show how some of our data is robust to reasonable variations in some of the chosen parameters. Specifically, we examine how the respective results are affected by the choice of stripe-area thresholds for calculating front speeds and survival probability rates, as well as changes in the fitting window for $\mathcal{P}_S$.

    %==================================================
    \begin{figure}
        \centering
        \includegraphics[width=0.85\linewidth]{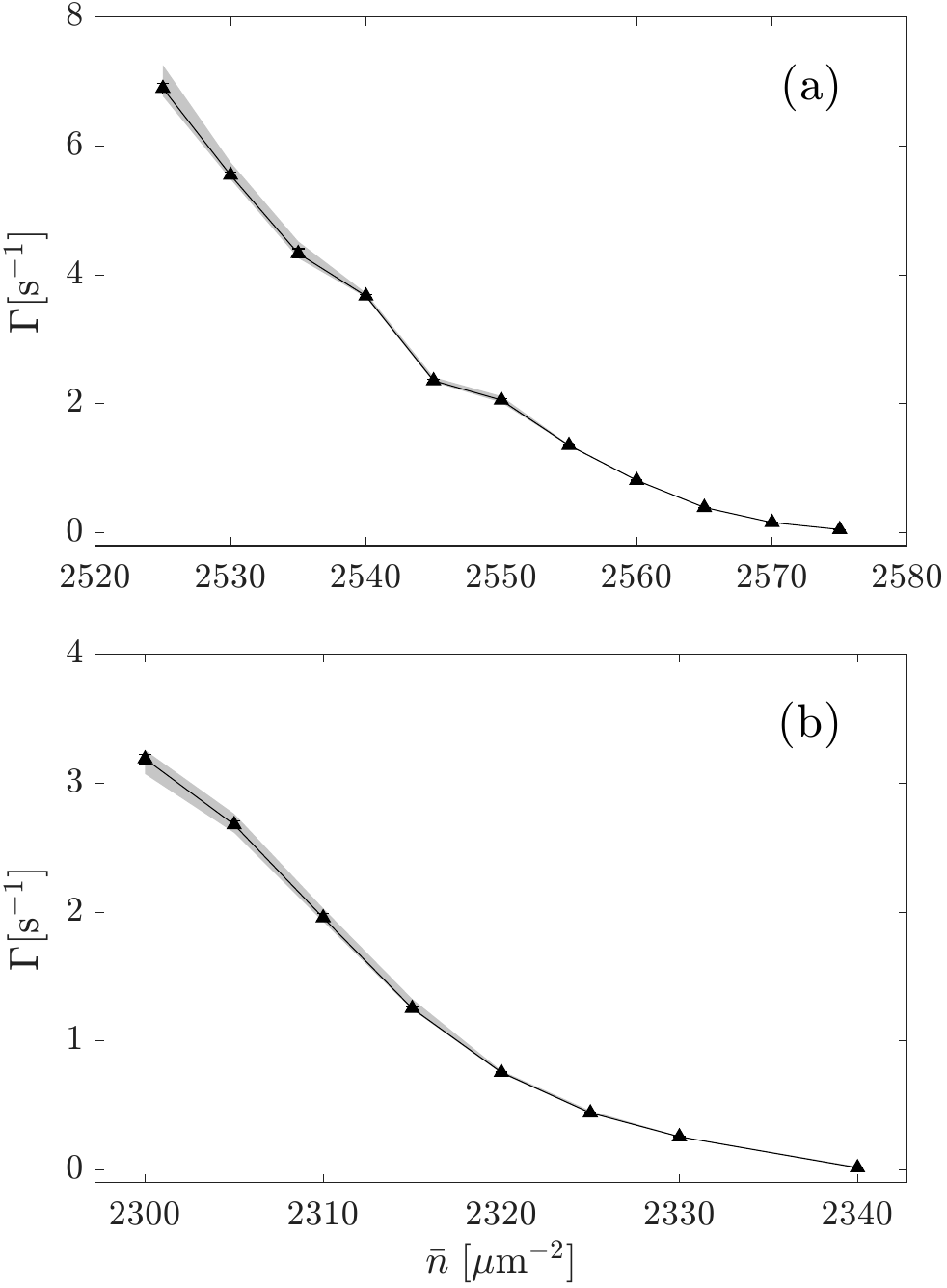}
        \caption{Stability of decay rates against the decay area threshold. The shaded plot area corresponds to the maximum variation in numerically-extracted rates as the survival probability is defined using a $1\%$ (upper bound) and $10\%$ (lower bound) stripe-area threshold. Similar to Fig.~\ref{fig:DecayRates} of the main text, scattering lengths are (a) $102.25\,a_0$  and (b) $102.50\,a_0$. Black triangles with error bars correspond to the same data of the $5\%$ threshold used in the main text.}
        \label{fig:Rate_vs_Athres}
    \end{figure}
	%==================================================
    Fig.~\ref{fig:Rate_vs_Athres} shows the numerically determined decay rates from SPeGPE simulations where the threshold for the onset of false-vacuum decay is defined by the decay area $\mathcal{A}(t)$ for each simulation continuously varied between $1\%$ (decay identified earlier, corresponding to maximum bound on $\Gamma$) and $10\%$ (slower identification, lower bound on $\Gamma$). Due to the relatively rapid growth of the bubbles after decay is triggered, varying the numerical threshold for identifying decay appears to have very little effect on the extracted decay rates.
    
    %==================================================
    \begin{figure}
        \centering
        \includegraphics[width=\linewidth]{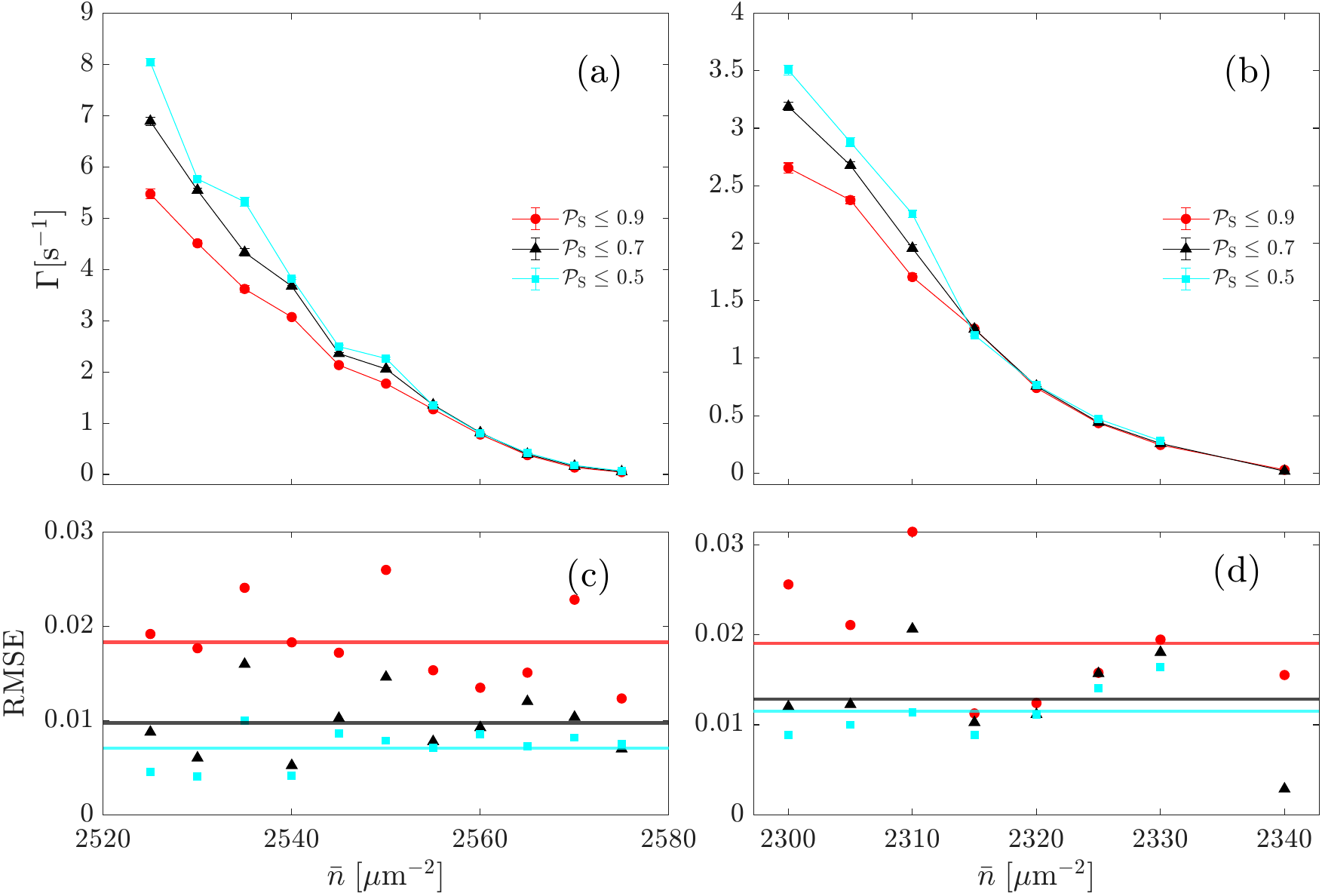}
        \caption{Variations in the decay rate by changing the survival probability fitting window. Upper row shows the numerical decay rates calculated for (a) $102.25\,a_0$  and (b) $102.50\,a_0$, using the fitting windows indicated. Lower row gives the root mean-squared error (RMSE) from Eq.~\eqref{eq:RMSE} for each curve above, with the mean RMSE shown as solid horizontal lines. Panels (b) and (d) are missing a data point at $\bar{n}=2340\,\mu\mathrm{m}^{-2}$, since the survival probability curve does not cross $50\%$ up to the maximum simulation time of $8\mathrm{s}$.}
        \label{fig:Rate_vs_Ps}
    \end{figure}
    
	%==================================================
    \begin{figure}
        \centering
        \includegraphics[width=\linewidth]{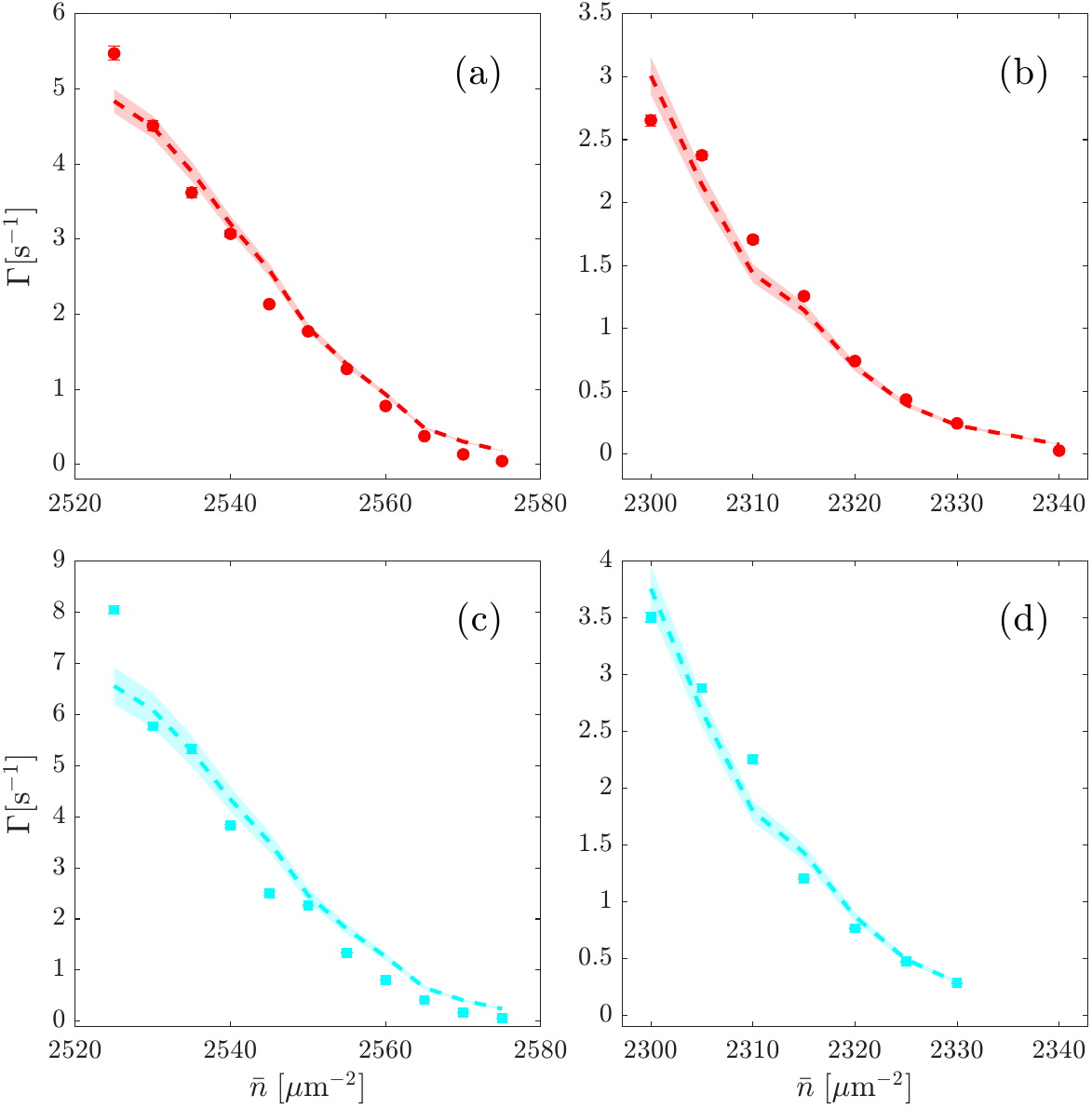}
        \caption{Rates of decay against the bounce solution, for the different $\mathcal{P}_\mathrm{S}$ windows shown in Fig.~\ref{fig:Rate_vs_Ps}. Panels (a) and (b) show the data for rates extracted with the wider window of $\mathcal{P}_\mathrm{S}\leq 0.9$, for $102.25\,a_0$  and $102.50\,a_0$, respectively. Lower panels (c) and (d) use instead a fitting window of $\mathcal{P}_\mathrm{S}\leq 0.5$ for the same scattering lengths as above. For each panel, the bounce solution ~\eqref{eq:GammaBounce} with the amplitude fit $A$ is shown as a blue dashed line, with the standard error on the fit shown as a shaded area, just as in Fig.~\ref{fig:DecayRates} of the main text.}
        \label{fig:Rate_Err_Bounce}
    \end{figure}
	%==================================================

    Fig.~\ref{fig:Rate_vs_Ps} shows the simulated decay as the fitting window for the survival probability $\mathcal{P}_{\mathrm{S}}$ is varied. Due to the fact that there is typically some delay after $t=0$ before typical exponential behaviour begins (i.e. $\mathcal{P}_{\mathrm{S}}(t)$ is either flat or at least sub-exponential), widening the window to $\mathcal{P}_{\mathrm{S}}\leq 0.9$ decreases the typical rate. This effect appears to persist past the value of 0.7 used in the main text and leads to an even higher rate of decay for $\mathcal{P}_{\mathrm{S}}\leq 0.5$. The variation is strongly density-dependent, and only results in a spread close to the shear instability (lower densities). The root mean-squared error (RMSE) is plotted in the lower panels. The RMSE is given by,
    \begin{equation}
        \mathrm{RMSE}=\sqrt{\frac{\sum_{t_i}|\mathcal{P}_{\mathrm{S}}(t_i)-A\mathrm{e}^{-\Gamma t_i}|^2}{N_t-2}}\;,
        \label{eq:RMSE}
    \end{equation}
    where the sum is taken over the discretized sample time points $t_i$ for a total of $N_t$ time steps in the fitting window, $A$ is the best-fit amplitude, and $\Gamma$ is the best-fit rate. The denominator $N_t-2$ corresponds to the number of residual degrees of freedom, since two parameters are fitted. For the data at $a_{\mathrm{s}}=102.50\,a_0$, there is no data point at $\bar{n}=2340\,\mu\mathrm{m}^{-2}$ for $\mathcal{P}_{\mathrm{S}}\leq 0.5$, since for this set of parameters, fewer than half the total 250 simulations decayed in the allotted simulation time.
    
    Fig.~\ref{fig:Rate_Err_Bounce} shows the same data as Fig.~\ref{fig:Rate_vs_Ps} (a)-(b), now with the corresponding instanton bounce calculations, i.e. the same format as Fig.~\ref{fig:DecayRates} of the main text, but corresponding amplitude fits $A$ to each data set. It becomes clear that although there is some variation in the absolute value of the decay rate derived from fitting the survival probability, the overall qualitative behaviour of the instanton bounce calculation is not strongly affected.

    \begin{figure}
        \centering
        \includegraphics[width=0.95\linewidth]{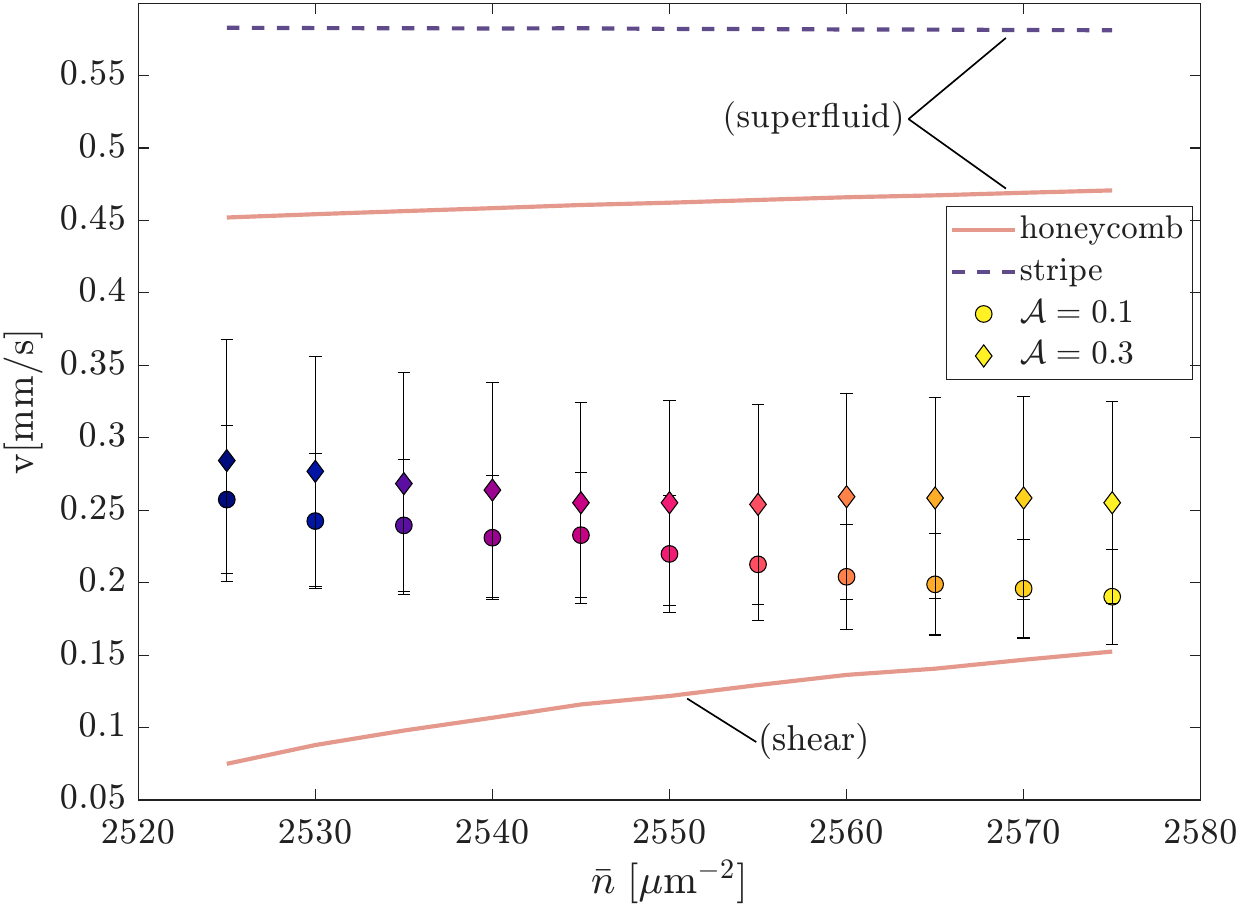}
        \caption{Similar to the inset Fig.~\ref{fig:SoundSpeeds}, now comparing against the bubble velocities calculated at $\mathcal{A}=0.3$ (diamonds), versus the choice of $\mathcal{A}=0.1$ (circles) from the main text.}
        \label{fig:VelocityErr}
    \end{figure}

    Fig.~\ref{fig:VelocityErr} shows the same comparison of bubble speeds as the inset of Fig.~\ref{fig:SoundSpeeds} in the main text, now comparing against speeds calculated at 30\% decayed area (i.e. at $\mathcal{A}(t^\star)=0.3$). While the mean velocities are slightly higher, the estimated error spread still places both results at approximately the same values. Furthermore, the overall trends are the similar, with the speed of bubble growth lying between the shear and superfluid speeds of sound,  and increasing as the shear instability is approached.
    
	%==============================================================================
    \section{Derivation of the bounce action}
    \label{app:BounceDerivation}
   Here, we derive the effective action and equations of motion in terms of the control parameters $c_i$. In the following, no assumptions are made about bubble shapes are made, but rather a structural rearrangement from a stripe state to a honeycomb state.
    Starting from the standard Euclidean extended Gross-Pitaevskii action, we have,
    \begin{align}
        S=&\;\int\mathrm{d}\tau\int\mathrm{d}\mathbf{x}\Bigl[\hbar\psi^*\partial_\tau\psi+\frac{\hbar^2}{2m}|\nabla\psi|^2+\frac{\tilde{g}}{2}|\psi|^4+\frac{2\tilde{\gamma}_Q}{5}|\psi|^5\nonumber\\&+\frac{\tilde{g}_{\mathrm{dd}}}{2}\int\mathrm{d}\mathbf{x}'U^{2\mathrm{D}}(\mathbf{x}-\mathbf{x}')|\psi(\mathbf{x}')|^2|\psi(\mathbf{x})|^2\Bigr]\,.
    \end{align}
    The real ansatz presented in Eq.~\eqref{eq:ansatz} forbids any dynamics since the Berry phase term becomes a total derivative. Hence, we shall write the wavefunction in Madelung form,
    \begin{equation}
        \psi(\mathbf{x},\tau)=\sqrt{\rho(\mathbf{x},\tau)}\,
        \mathrm{e}^{\mathrm{i}\theta(\mathbf{x},\tau)}\,,
    \end{equation}
    into the eGP action, which, using the fact that $\int\mathrm{d}\mathbf{x}\;\partial_\tau \rho=0$, gives the density-phase representation of the action,
    \begin{align}
        S=&\int\mathrm{d}\tau\mathrm{d}\mathbf{x}\Big\{
        \mathrm{i}\hbar \rho\partial_\tau \theta
        +\frac{\hbar^2}{2m}\left[(\nabla\sqrt{\rho})^2+\rho(\nabla\theta)^2\right]+\frac{\tilde{g}}{2}\rho^2\nonumber\\&+\frac{2\tilde{\gamma}_Q}{5}\rho^{\frac{5}{2}}+\frac{\tilde{g}_{\mathrm{dd}}}{2}\int\mathrm{d}\mathbf{x}'U^{2\mathrm{D}}(\mathbf{x}-\mathbf{x}')\rho(\mathbf{x}')^2\rho(\mathbf{x})^2
        \Big\}\;.\label{eq:GPActionRho}
    \end{align}
    
    Let us now suppose,
    \begin{equation}
        \theta(\mathbf{x},\tau)=\sum_{j=1}^{\mathcal{N}}\dot{c}_j(\tau)\chi_j(\mathbf{x},\mathbf{c})\;,
        \label{eq:thetaofcandchi}
    \end{equation}
    where the unknown function $\chi_j$ is a function of time-dependent real-valued amplitudes $\mathbf{c}=(c_1,c_2,...,c_\mathcal{N})$, which parametrize, e.g., the real-valued field Eq.~\eqref{eq:ansatz} and thus the density $\rho$. The logic of the choice Eq.~\eqref{eq:thetaofcandchi} is as follows: a nonzero flow of the amplitudes $\mathbf{c}$ in time $\tau$ corresponds to a nonzero change in the time-dependent amplitudes of the field $\psi$, and therefore the phase must depend at least on the corresponding time dependence of the density. 
    
    Variation of $S$ with respect to $\rho$ leads to the imaginary-time analogue of the classical continuity equation,
    \begin{equation}
       0=\mathrm{i}\hbar\partial_\tau\rho+\frac{\hbar^2}{m}\nabla\cdot(\rho\nabla\theta)
    \end{equation}
    and thus, for arbitrary infinitesimal changes of $\mathbf{c}$, to the defining equations for $\chi_j$,
    \begin{equation}
        \mathrm{i}\frac{\partial \rho}{\partial c_j}+\frac{\hbar}{m}\nabla\cdot\left[\rho\nabla\chi_j\right]=0
        \,,\quad j=1,...,\mathcal{N}\;.~\label{eq:EllipticChi}
    \end{equation}
    Multiplying by $\chi_i$ and integrating (by parts for the gradient term) over the unit cell yields,
    % integrating,
    % \begin{equation}
    %     \int\mathrm{d}\mathbf{x}\chi_i\frac{\partial \rho}{\partial c_j}=-\frac{\hbar}{m}\int\mathrm{d}\mathbf{x}\chi_i\nabla\cdot\left[\rho\nabla\chi_j\right]\;,
    % \end{equation}
    \begin{align}
        \mathrm{i}\!\int\mathrm{d}\mathbf{x}\,\chi_i\frac{\partial \rho}{\partial c_j}
        =-\frac{\hbar}{m}\bigg\{
        &\int_{\partial\Omega}\chi_i\rho\nabla\chi_j\cdot\hat{\mathbf{n}}\mathrm{d}s
        -\int\mathrm{d}\mathbf{x}\,\rho\nabla\chi_i\cdot\nabla\chi_j\bigg\},
    \end{align}
    where here $\partial\Omega$ is the boundary of our integration volume $\Omega$, d$s$ the scalar measure on the boundary, and $\hat{\mathbf{n}}$ is a unit vector pointing normal to the border of the cell. Due to the periodicity of the system, this boundary term must vanish. Hence, we obtain,
    \begin{equation}
        \mathrm{i}\!\int\mathrm{d}\mathbf{x}\,\chi_i\frac{\partial \rho}{\partial c_j}=\frac{\hbar}{m}\int\mathrm{d}\mathbf{x}\,\rho\nabla\chi_i\cdot\nabla\chi_j\;.\label{eq:ChiRelation}
    \end{equation}
    
    The first term of the Lagrangian density can be re-expressed as follows,
    \begin{align}
        \mathrm{i}\!\int\!\mathrm{d}\tau\mathrm{d}\mathbf{x}\,\hbar\rho\partial_\tau\theta
        =&\int\!\mathrm{d}\tau\mathrm{d}\mathbf{x}\left(\hbar\rho\sum_j\ddot{c}_j\chi_j+\hbar\rho\sum_{ij}\dot{c}_j\dot{c}_i\frac{\partial \chi_j}{\partial c_i}\right)\\
        =&\;-\int\mathrm{d}\tau\mathrm{d}\mathbf{x}\;\sum_{ij}\hbar\dot{c}_j\dot{c}_i\chi_j\frac{\partial\rho}{\partial c_i}\,,
        \label{eq:EuclidFirstTerm}
    \end{align}
    where we have integrated the first term on the r.h.s.~by parts with respect to $\tau$, eliminating the boundary term with $\dot{c}_j(\tau\to0)=\dot{c}_j(\tau\to\infty)=0$ (see discussion below), and used the fact that, in the flow, $\partial_\tau\rho=\sum_i\dot{c}_i\partial \rho/\partial c_i$. 
    
    Next, we note that the classical kinetic-energy density of the flow reads:
    \begin{equation}
        \frac{\hbar^2}{2m}\rho(\nabla\theta)^2=\frac{\hbar^2}{2m}\rho\sum_{ij}\dot{c}_j\dot{c}_i\nabla\chi_i\cdot\nabla \chi_j\;.\label{eq:MadelungKE}
    \end{equation}
    Inserting Eq.~\eqref{eq:ChiRelation} into Eq.~\eqref{eq:EuclidFirstTerm}, and simplifying with Eq.~\eqref{eq:MadelungKE}, the action takes its final form,
    \begin{equation}
        S=\int\mathrm{d}\tau\left[-\frac{1}{2}\sum_{ij}\dot{c}_iM_{ij}(\mathbf{c})\dot{c}_j+E_\rho(\mathbf{c})\right]\,,
        \label{eq:Seff_wint}
    \end{equation}
    with metric tensor,
    \begin{equation}
        M_{ij}(\mathbf{c})=\frac{\hbar^2}{m}\int\mathrm{d}\mathbf{x}\;\rho\nabla\chi_i\cdot\nabla \chi_j\,,
    \end{equation}
    and the eGP energy of the density variations, i.e., for constant phase,
    \begin{equation}
        E_{\mathrm{eGP}}[\rho,\theta=\mathrm{const.}]\equiv
        E_{\rho}(\mathbf{c})=E_K(\mathbf{c})+E_{\mathrm{int}}(\mathbf{c})\,,
        \label{eq:Edensity}
    \end{equation}
    including the kinetic energy of the density variations,
    \begin{equation}
        E_{K}(\mathbf{c})=\frac{\hbar^2}{2m}\int\mathrm{d}\mathbf{x}\;(\nabla\sqrt{\rho})^2\,,
        \label{eq:EKdensity}
    \end{equation}
    and the interaction energy,
    \begin{align}
        E_{\mathrm{int}}(\mathbf{c})=&\;\int\mathrm{d}\mathbf{x}\;\biggl[\frac{\tilde{g}}{2}\rho(\mathbf{x})^2+\frac{2\tilde{\gamma}_Q}{5}\rho(\mathbf{x})^{\frac{5}{2}}\nonumber\\&+\frac{\tilde{g}_{\mathrm{dd}}}{2}\int\mathrm{d}\mathbf{x}'\rho(\mathbf{x})U^{2\mathrm{D}}(\mathbf{x}-\mathbf{x}')\rho(\mathbf{x}')\biggr]\,.
    \end{align}
    In order to calculate the action, the metric tensor must be calculated by solving the elliptic equation \eqref{eq:EllipticChi} at each point in $\mathbf{c}$-space. 

    The bounce can be determined by the variational minimization of Eq.~\eqref{eq:Seff_wint}, leading to the equations of motion,
    \begin{equation}
        \sum_{j}M_{ij}\ddot{c}_j+\sum_{jk}\Gamma_{ijk}\dot{c}_j\dot{c}_k+\frac{\partial E_{\rho}}{\partial c_i}=0\;,
    \end{equation}
    where 
    \begin{equation}
        \Gamma_{ijk}=\frac{1}{2}\left(\frac{\partial M_{ij}}{\partial c_k}+\frac{\partial M_{ik}}{\partial c_j}-\frac{\partial M_{jk}}{\partial c_i}\right)
    \end{equation}
    are the lower-index Christoffel symbols.

    One can show that the energy,
    \begin{equation}
        \tilde{\mathcal{E}}= \sum_i\dot{c}_i\frac{\partial\mathcal{L}}{\partial\dot{c}_i}-\mathcal{L}
    \end{equation}
    is conserved along the classical path by verifying $\mathrm{d}\tilde{\mathcal{E}}/\mathrm{d}\tau=0$, with,
    \begin{equation}
        \mathcal{L}=-\frac{1}{2}\sum_{jk}\dot{c}_j\dot{c}_kM_{jk}+E_\rho(\mathbf{c})\;.
    \end{equation}

    In the case of a localized bounce trajectory, we have $\dot{\bar{\mathbf{c}}}_\mathrm{F}=0$, which was also used in the derivation of $S$ to eliminate a boundary term. This is only strictly true in the $\beta\to\infty$ limit, but for sufficiently small temperatures, $\dot{\bar{\mathbf{c}}}$ must vanish at the turning point such that one can still consider a dilute bounce gas~\cite{Coleman1977FateI,Altland2010Book}. In this case, we have the `quantum' condition,
    \begin{equation}
        \mathcal{E}(\mathbf{c}_\mathrm{F})=-E_\rho(\mathbf{c}_\mathrm{F})\;,
    \end{equation}
    leading to the fact that, along the path,
    \begin{equation}
        -E_\rho(\mathbf{c}_\mathrm{F})=-\frac{1}{2}\sum_{jk}\dot{\bar{c}}_j\dot{\bar{c}}_kM_{jk}-E_{\rho}(\bar{\mathbf{c}})\;.
    \end{equation}
    Noting that one can write the first fundamental form in this space as,
    \begin{equation}
        (\mathrm{d}\mathbf{s})^2=\sum_{ij}\mathrm{d}c_iM_{ij}\mathrm{d}c_j\,,
    \end{equation}
    allows us to write,
    \begin{equation}
        \mathrm{d}\tau=\sqrt{\frac{-(\mathrm{d}\mathbf{s})^2}{2\left[E_{\rho}(\bar{\mathbf{c}})-E_\rho(\mathbf{c}_\mathrm{F})\right]}}\,.
    \end{equation}
    Since $M_{ij}<0$ in the imaginary-time formulation,  $(\mathrm{d}\mathbf{s})^2<0$ needs to hold along the path. It is more convenient conceptually to work in a space with a positive metric, so taking $\tilde{M}_{ij}=-M_{ij}$ with corresponding $(\mathrm{d}\tilde{\mathbf{s}})^2=-(\mathrm{d}\mathbf{s})^2$, which (dropping the tildes) gives Eq.~\eqref{eq:BounceE}.

	%==============================================================================
    \section{2D supersolid ansatz}
    \label{app:ParamSpace}
    
    In Appendix \ref{app:BounceDerivation}, no assumptions about the number of parameters in $\mathbf{c}$ were made. Now, we shall assume that the parameter space is two-dimensional via our ansatz,
    \begin{align}
        \psi(x,y)=\sqrt{\bar{n}}\Bigg[
        &c_0
        -c_1\cos\left(\frac{2ky}{\sqrt{3}}\right)
        -2c_2\cos\left(\frac{ky}{\sqrt{3}}\right)\cos(kx)\Bigg]\,,
        \label{eq:app:ansatz}
    \end{align}
    where $k=2\pi/L_{\mathrm{uc}}$, $\bar{n}$ is the mean density, and $L_{\mathrm{uc}}$ is the side length of a single rhombus unit cell. This choice comes from considering the expansion,
   \begin{align*}
        \psi(x,y)=\sqrt{\bar{n}}\sum_{i}c_i\cos(\mathbf{k}_i\cdot \mathbf{x})\,,
	\end{align*}
    with wavevectors 
    \begin{align}
        \mathbf{k}_i\in&{\Bigl\{(0,0),
        \left(0,\pm 2k/\sqrt{3}\right)},{\small\left(\pm k,\pm k/\sqrt{3}\right),
        \left(\pm k,\mp k/\sqrt{3}\right)\Bigr\}}\,,
    \end{align}
    and simplifying under the assumption that the amplitudes of $\pm(k,k/\sqrt{3})$ and $\pm(k,-k/\sqrt{3})$ are the same. Extending our ansatz further implies loosening this assumption, or going farther by taking into account higher harmonics with similar symmetry. 
    
    The integral of the density over the unit cell area, 
    \begin{align}
    A_{\mathrm{uc}}=\int_{\mathrm{uc}}\mathrm{d}\mathbf{x}
    &=\int_0^{\sqrt{3}L_{\mathrm{uc}}/2}
    \left(\int_{y/\sqrt{3}}^{y/\sqrt{3}+L_{\mathrm{uc}}}\mathrm{d}x\right)\mathrm{d}y
    \nonumber\\
    &={\sqrt{3} L_{\mathrm{uc}}^2}/{2}\,, 
    \end{align}
    gives
    \begin{align}
        N_{\mathrm{uc}}\equiv \int_{\mathrm{uc}}\mathrm{d}\mathbf{x}\;|\psi|^2\,,
    \end{align}
    and since $N_{\mathrm{uc}}=\bar{n}A_{\mathrm{uc}}$, choosing the constant
    \begin{equation}
        c_0=\sqrt{1-c_1^2/2-c_2^2}
    \end{equation}
    ensures normalization of $\psi$. The eGP energy of the density variations can be split into  different contributions,
    \begin{equation}
        E_{\rho}(\mathbf{c})=E_K+E_c+E_Q+E_D\,,
        \label{eq:GPEnergy}
    \end{equation}
    corresponding to the kinetic, contact, quantum-fluctuation, and dipolar energies, respectively.%. such that $E_{\mathrm{int}}=E_c+E_Q+E_D$.

    The kinetic energy contribution can straightforwardly be calculated,
    \begin{align}
        E_K(\mathbf{c})=&\;\frac{\hbar^2}{2m}\int\mathrm{d}\mathbf{x}\;(\nabla\sqrt{\rho})^2=\frac{\hbar^2}{2m}\frac{4 \pi ^2 \bar{n}}{\sqrt{3}} \left(c_1^2+2 c_2^2\right)
    \end{align}
    The contact and quantum fluctuation terms are, respectively,
    \begin{widetext}
    \begin{align}
        E_c(c_1,c_2)=\frac{\tilde{g}}{2}\int\mathrm{d}\mathbf{x}|\psi|^4=&-\frac{\tilde{g}}{2}\frac{1}{16} \sqrt{3} L^2 \bar{n}^2 \left(7 c_1^4+16 c_1^2 \left(c_2^2-1\right)+48\, c_1 c_2^2 \sqrt{1-c_1^2/2-c_2^2}+22 c_2^4-32 c_2^2-8\right)\,,\\
        E_{Q}(c_1,c_2)= \frac{2}{5}\tilde{\gamma}_{Q}\int\mathrm{d}\mathbf{x}|\psi|^5=&\;
            \frac{1}{5}\tilde{\gamma}_{Q}\frac{1}{8}\sqrt{3}\,L^{2} \bar{n}^{5/2}\Bigl(
                  30\,c_{1}^{3}c_{2}^{2}
                + 16\,c_{1}^{2}\!\left(3c_{2}^{2}+2\right)
                    \sqrt{1-c_{1}^{2}/2-c_{2}^{2}}
                \nonumber\\
              &\;+2\,\left(9c_{2}^{4}+32c_{2}^{2}+4\right)
                    \sqrt{1-c_{1}^{2}/2-c_{2}^{2}}  
                    - 3\,c_{1}^{4}\sqrt{1-c_{1}^{2}/2-c_{2}^{2}}
                + 60\,c_{1}c_{2}^{2}\!\left(c_{2}^{2}-2\right)
            \Bigr)\,.
    \end{align}
    \end{widetext}
    The energy contribution due to the DDI is more complicated,
    \begin{equation}
        E_{D}\equiv\frac{\tilde{g}_{\mathrm{dd}}}{2}\int\mathrm{d}\mathbf{x}\int\mathrm{d}\mathbf{x}'|\psi(\mathbf{x})|^2 |\psi(\mathbf{x}')|^2 U^{2D}(\mathbf{x}-\mathbf{x}')\,.
    \end{equation}
    For our ansatz Eq.~\eqref{eq:ansatz}, the Fourier transform of the density can be calculated exactly, and corresponds to a sum of delta functions weighted by simple functions of the amplitudes,
       \begin{equation}
        \tilde{\rho}(\mathbf{k})=\bar{n}\sum_{i}K_i(c_1,c_2)\delta(k_x-k_{i,x})\delta(k_y-k_{i,y})\,,
        \label{eq:DensityMomentumExpansion}
    \end{equation}
    where the sum runs over a particular finite set of 19 momentum-space points $\mathbf{k}=\mathrm{n}\mathbf{b}_1 
    +\mathrm{m}\mathbf{b}_2$ with (reciprocal) lattice vectors $\mathbf{b}_1=({2\pi}/{L_{\mathrm{uc}}})(1,1/\sqrt{3})$ and $\mathbf{b}_2=({4\pi}/{L_{\mathrm{uc}}})(0,1/\sqrt{3})$, and subject to $\mathrm{max}(|\mathrm{n}|,|\mathrm{m}|,|\mathrm{m}+\mathrm{n}|)\leq 2$. In simpler terms, these points correspond to $\mathbf{k}=\mathbf{0}$ and the two surrounding hexagons are constructed via the lattice vectors $\mathbf{b}_i$. The dipole-dipole energy thus results as
    \begin{equation}
        E_{D}=\frac{\tilde{g}_{\mathrm{dd}}}{2}\sum_{i}|K_i(c_1,c_2)|^2\tilde{U}^{2D}(\mathbf{k}_{i})\,.
    \end{equation}
    The coefficients can each be exactly calculated and are shown in Table \ref{tab:KCoeff}.

    \begin{table}[t]
    \centering
    \begin{tabular}{|ccc|}
        \hline $\dfrac{L|k_x|}{2\pi}$ &
        $\dfrac{\sqrt{3}L|k_y|}{2\pi}$ &
        $K_i/\pi^2$ \\\hline
        0 & 0 & $4$ \\
        0 & 2 & $2c_2^2 - 2\sqrt{2}\,c_1 \sqrt{2-c_1^2-2c_2^2}$ \\
        0 & 4 & $c_1^2$ \\
        1 & 1 & $2c_1c_2 - 2\sqrt{2}\,c_2 \sqrt{2-c_1^2-2c_2^2}$ \\
        1 & 3 & $2c_1c_2$ \\
        2 & 0 & $2c_2^2$ \\
        2 & 2 & $c_2^2$ \\
        \hline
    \end{tabular}
    \caption{Coefficients $K_i$ in Eq.~\eqref{eq:DensityMomentumExpansion}. Each row applies to the set of all combinations of sign choices of $k_x$ and $k_y$. All other momentum contributions are zero.}
    \label{tab:KCoeff}
\end{table}
	%==================================================
    The classical equations of motion for the bounce action are then
    \begin{widetext}
        \begin{align}
         \ddot{c}_1M_{11}+\ddot{c}_2M_{12}=&\;-\frac{1}{2}\left[\dot{c}_1^2\frac{\partial M_{11}}{\partial c_1}-\dot{c}_2^2\frac{\partial M_{22}}{\partial c_1}\right]-\left[\dot{c}_1\dot{c}_2\frac{\partial M_{11}}{\partial c_2}+\dot{c}_2^2\frac{\partial M_{12}}{\partial c_2}\right]+\frac{\partial E_{\rho}}{\partial c_1}\,,\label{eq:Eqddot1}\\
         \ddot{c}_1M_{12}+\ddot{c}_2M_{22}=&\;\frac{1}{2}\left[\dot{c}_1^2\frac{\partial M_{11}}{\partial c_2}-\dot{c}_2^2\frac{\partial M_{22}}{\partial c_2}\right]-\left[\dot{c}_1\dot{c}_2\frac{\partial M_{22}}{\partial c_1}+\dot{c}_1^2\frac{\partial M_{12}}{\partial c_1}\right]+\frac{\partial E_{\rho}}{\partial c_2}\,.\label{eq:Eqddot2}
    \end{align}
   \end{widetext}
 
    Equations \eqref{eq:Eqddot1}-\eqref{eq:Eqddot2} can be reduced to a set of four coupled first-order nonlinear equations with fixed boundary conditions $\dot{\mathbf{c}}(\tau \to0)=\dot{\mathbf{c}}(\tau \to\infty)=0$, and $\mathbf{c}(\tau \to+\infty)=\mathbf{c}_{\text{H}}$. Since the path is symmetric, it is numerically simpler to solve the equations by starting at the turning point, and finding the path that returns to $\mathbf{c}_{\mathrm{F}}$ (e.g. using the shooting method) since it must lie on the equipotential line of $E_{\rho}(\mathbf{c}_{\mathrm{F}})$.

    %==============================================================================
    \section{Effects of increased temperature}
    \label{app:Temperature}

    In the main text, we consider FVD simulations using the SPeGPE [Eq.~\eqref{eq:SPGPE}], at $T=5\,\mathrm{nK}$, and we find that our minimal model of a bounce solution fits the simulated decay rates reasonably well. The expressions given in Eqs.~\eqref{eq:GammaBounce} and \eqref{eq:B1} correspond to those in the quantum tunneling theory originally developed by Coleman. At high temperatures, the kinetic-energy contribution can be neglected, and so the inverse temperature $\beta$ becomes a prefactor to the difference in action \cite{Linde1983Decay},
    \begin{equation}
        B\to B_T
        =\beta\left[E_{\rho}(\mathbf{c}_\mathrm{saddle})-E_{\rho}(\mathbf{c}_\mathrm{F})\right]\;.
    \end{equation}
    Hence, the rate becomes 
    \begin{equation}
        \Gamma_T=A B_T\,\mathrm{e}^{-B_T/\hbar}\;.
        \label{eq:GammaT}
    \end{equation}
    We shall refer to the bounce action and corresponding decay rates calculated with $B$ and $\Gamma$ as `quantum bounce', and those with $B_T$ and $\Gamma_T$ as `thermal'.

    \begin{figure}[hb]
        \centering
        \includegraphics[width=1.0\linewidth]{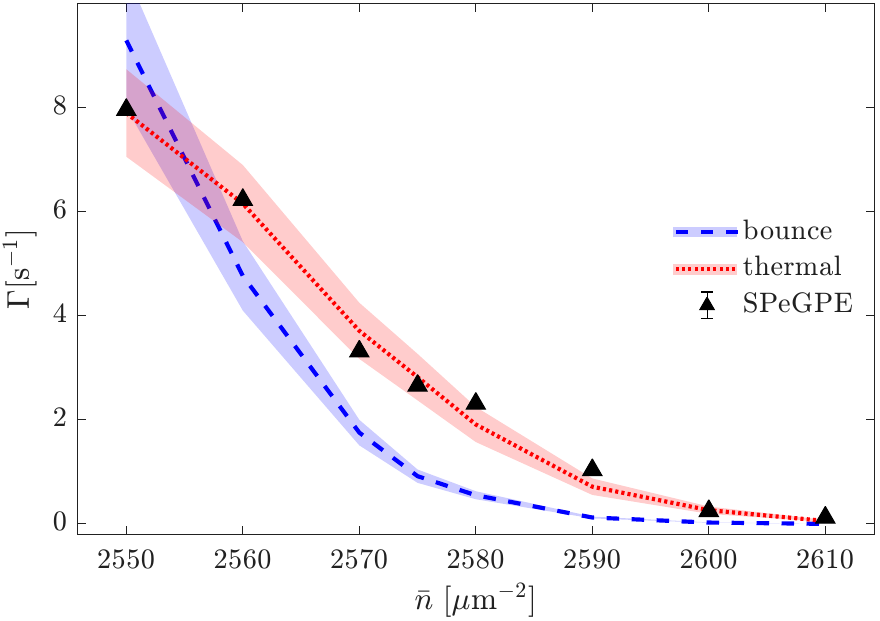}
        \caption{Similar to Fig.~\ref{fig:DecayRates} (a), showing the vacuum decay rate as a function of mean density at $102.25\,a_0$, now with simulations performed at $T=10\,\mathrm{nK}$. The single-parameter amplitude bounce fit is plotted in blue as before, now with the two-parameter thermal fit from Eq.~\eqref{eq:GammaT} shown in red.}
        \label{fig:FiniteT}
    \end{figure}

    In Fig.~\ref{fig:FiniteT}, we show simulations similar to Fig.~\ref{fig:DecayRates} (a) in the main text, but at a higher temperature of $T=10\,\mathrm{nK}$. We immediately see that the quantum bounce theory with only the single amplitude fit fails to accurately capture the correct trend of the decay rate. To compare with the thermal case, we follow the approach of Ref.~\cite{Zenesini2024False}, and plot Eq.~\eqref{eq:GammaT} by fitting both the amplitude $A$ and the inverse temperature $\beta$ (see also \cite{Sivasankar2025Temperature}, where the action $B_T$ was fit instead). The thermal approach leaves us with a better fit to the data, and so we estimate that in this region, the decay is dominated by temperature-seeded fluctuations. We find that the effective temperature via the fit corresponds to $(k_B\beta_{\mathrm{fit}})^{-1}=1.4\times10^{-2}\mathrm{nK}$. We can compare this to the data in Fig.~\ref{fig:DecayRates} of the main text, which both give a lower effective temperature  of approximately $(k_B\beta_{\mathrm{fit}})^{-1}=6\times10^{-3}\mathrm{nK}$ (thermal model not shown in the main text). It is possible that a better comparison with the barrier height in Sec.~\ref{subsec:effectiveaction} can be made with this extracted effective temperature, however one must still take care that the two-parameter fit does not capture all of the physics of the full theory. Since intermediate morphologies might not be well-captured by the two-parameter theory, a more reliable comparison might be able to be made by extending the $c_j$ parameters to higher harmonics.
    \end{appendix}
\end{document}